\PassOptionsToPackage{a4paper,margin=24mm}{geometry}   
\PassOptionsToPackage{hidelinks,breaklinks}{hyperref}  

\documentclass[preprint,1p]{elsarticle}

\usepackage[T1]{fontenc}
\usepackage{lmodern}
\usepackage{amsmath,amssymb}
\usepackage{graphicx}
\usepackage{booktabs}
\usepackage{siunitx}
\usepackage{xurl}        
\usepackage{microtype}
\usepackage{geometry}    
\usepackage{listings}  
\usepackage{hyperref}    
\usepackage{longtable}
\usepackage{orcidlink}

\usepackage{xspace}
\DeclareRobustCommand{\pkg}[1]{\texttt{#1}}
\DeclareRobustCommand{\SVEMnet}{\pkg{SVEMnet}\xspace}
\DeclareRobustCommand{\glmnet}{\pkg{glmnet}\xspace}
\DeclareRobustCommand{\wAIC}{\pkg{wAIC}\xspace}
\DeclareRobustCommand{\wBIC}{\pkg{wBIC}\xspace}
\DeclareRobustCommand{\wSSE}{\pkg{wSSE}\xspace}
\DeclareRobustCommand{\pfull}{\ensuremath{p_{\mathrm{full}}}\xspace}

\lstdefinestyle{rcompact}{
  language=R,
  basicstyle=\ttfamily\footnotesize,
  columns=fullflexible,
  keepspaces=true,
  showstringspaces=false,
  upquote=true,
  tabsize=2,
  breaklines=true,
  breakatwhitespace=true,
  numbers=left,
  numberstyle=\tiny,
  numbersep=4pt,
  aboveskip=3pt,
  belowskip=3pt,
  xleftmargin=0pt,
  frame=none
}
\date{22 October 2026}
\journal{Chemometrics and Intelligent Laboratory Systems}

\begin{document}

\begin{frontmatter}
\title{SVEMnet: An R package for Self-Validated Elastic-Net Ensembles and Multi-Response Optimization in Small-Sample Mixture--Process Experiments}

\author[inst1]{Andrew T. Karl\orcidlink{0000-0002-5933-8706}}
\cortext[cor1]{Corresponding author}
\ead{akarl@asu.edu}
\affiliation[inst1]{
  organization={Karl Statistical Services},
  city={Aurora},
  state={CO},
  postcode={80016},
  country={USA}
}

\begin{abstract}
\SVEMnet is an R package for fitting Self-Validated Ensemble Models (SVEM) with elastic-net base learners and performing multi-response optimization in small-sample mixture--process design-of-experiments (DOE) studies with numeric, categorical, and mixture factors. \SVEMnet wraps elastic-net and relaxed elastic-net models for Gaussian and binomial responses from \glmnet{} in a fractional random-weight (FRW) resampling scheme with anti-correlated train/validation weights; penalties are selected by validation-weighted AIC- and BIC-type criteria, and predictions are averaged across replicates to stabilize fits near the interpolation boundary. In addition to the core SVEM engine, the package provides deterministic high-order formula expansion, a permutation-based whole-model test heuristic, and a mixture-constrained random-search optimizer that combines Derringer--Suich desirability functions, bootstrap-based uncertainty summaries, and optional mean-level specification-limit probabilities to generate scored candidate tables and diverse exploitation and exploration medoids for sequential fit--score--run--refit workflows. A simulated lipid nanoparticle (LNP) formulation study illustrates these tools, and simulation experiments based on sparse quadratic response surfaces benchmark \SVEMnet against repeated cross-validated elastic-net baselines.
\end{abstract}


\end{frontmatter}

\section{Introduction}

Formulation optimization experiments in chemistry and biopharmaceutics often
operate with limited runs, complex factor spaces (interactions, curvature,
mixtures), and multiple responses of interest. In such small-$N$ regimes,
flexible models can be unstable: small changes in the construction of the training sample (for
example, different train/validation splits or cross-validation folds) may yield
noticeably different predictions and selected penalties										 
\citep{breiman1996bagging}. Self-Validated Ensemble Models (SVEM) were proposed
by \citet{Lemkus2021} to mitigate this instability by
drawing fractional random-weight (FRW) \citep{Xu2020_FRW} bootstraps paired with anti-correlated
validation weights and averaging predictions across resampled fits. The idea was
presented under the name ``autovalidation'' by \citet{Gotwalt2018} and has
since been extended with a permutation-based whole-model testing heuristic
\citep{Karl2024_svem_test} and nonlinear base learners such as neural networks
in mixture experiments \citep{RamseyGaudardLevin2021_MixtureSVEM}. SVEM is primarily intended as a predictive surrogate-modeling tool rather than a hypothesis-testing procedure, in line with the broader
``to explain or to predict'' distinction of \citet{Shmueli2010ExplainPredict}.

SVEM has already been implemented in JMP Pro (with forward-selection and Lasso base learners) and a proprietary JMP add-in (using neural network base learners, \citep{RamseyGaudardLevin2021_MixtureSVEM}). However, many 
analysts work in R, and reproducible small-sample workflows benefit from an
open-source implementation that can be scripted, versioned, and embedded in simulation
and optimization pipelines. To this end, this paper introduces \SVEMnet, an R package that implements SVEM with elastic-net and optional relaxed elastic-net base learners
\citep{en05,Friedman2010_glmnet,relax07,Friedman2025_relax}. In a relaxed fit,
\glmnet{} computes a penalized solution path indexed by the penalty parameter
$\lambda$, identifies the active set at each $\lambda$, and then refits an unpenalized linear or logistic model on that active set, interpolating between
penalized and unpenalized coefficients via a relaxation parameter
$\gamma\in[0,1]$. In the sparse Gaussian linear model studied by \citet{relax07}, the relaxed lasso reduces the shrinkage of non-zero coefficients and can achieve lower prediction error than the standard lasso. Simulations in Section~\ref{sec:sim} indicate that combining relaxed base
learners with the SVEM FRW ensemble can further reduce out-of-sample
prediction error for Gaussian responses, plausibly because the ensemble averages over the additional
variance introduced by relaxation.
In place of the validation-weighted sum of squared errors criterion (denoted
\wSSE) used in earlier SVEM work by \citet{Lemkus2021}, \SVEMnet standardizes on
heuristic, validation-weighted information-criterion analogs---a validation-weighted Akaike information criterion (AIC) analog (\wAIC{}) and a validation-weighted Bayesian information criterion (BIC; Schwarz criterion) analog (\wBIC{}), with \wAIC{} the default for Gaussian responses and \wBIC{} the default for binomial responses
\citep{Akaike1974,Schwarz1978,BurnhamAnderson2002}. Flexible fits such as the lasso are known to behave poorly near the
interpolation boundary, where the number of runs $n$ approaches the full
expansion dimension $\pfull$ (the total number of coefficients in the ``full'' linear model
expansion, including the intercept), with ``peaking'' in out-of-sample error
documented in small-sample classification and regression problems
\citep{RAUDYS1998385} and in lasso
model selection \citep{Kraemer2009LassoPeaking}. Related analyses of the
$p\gtrsim n$ interpolation setting in linear prediction underscore that
performance can shift sharply near this boundary \citep{Hastie2022Ridgeless,Christensen2025}.
In the simulations reported
in Section~\ref{sec:sim}, \wAIC{} and \wBIC{} avoid this peaking, whereas
the loss-only \wSSE{} selector exhibits a pronounced spike at
$n\approx\pfull$.

Self-Validated Ensemble Models have already seen practical use in industrial DOE and chemometric settings, including mixture--process workflows for lipid nanoparticle (LNP) formulations and other pharmaceutical and analytical applications \citep{JoVE_LNP,KarlEtAl2022_JMPMixture,RamseyGaudardLevin2021_MixtureSVEM,RamseyGotwalt2018_JMPEurope,RamseyEtAl2021_SVEMParadigm,RamseyMcNeill2023_CMC,Mirzaiyanrajeh2023,Korany2024,Mostafa2023a,Hanafy}. These
applications motivate an implementation tailored to chemometric DOE. In
particular, \SVEMnet{} is designed to automate many of the steps in the SVEM-based
mixture--process workflow for LNP formulation optimization described by
\citet{JoVE_LNP}. \SVEMnet supplies an open-source SVEM implementation
over \glmnet{} with validation-weighted AIC/BIC selectors, deterministic
high-order full model expansions, a permutation-based whole-model test
heuristic, and a mixture-constrained random-search scoring tool for
multi-response desirability and candidate generation.

\SVEMnet is intended for round-by-round iteration in small-$N$
DOE: fit SVEM models on the current data; use the random-search optimizer to
generate and score a feasible candidate set under mixture and process
constraints; select a small number of high-score exploitation medoids and
high-uncertainty exploration medoids; then run those settings, append the new
data, and refit the models. The LNP formulation example below illustrates this workflow
\citep{JoVE_LNP,Karl2024_svem_test}. The sequential workflow is an iterate--append loop that alternates
exploitation and exploration based on predictive uncertainty, rather than a full Gaussian-process-based Bayesian optimization scheme, which remains the standard choice when richer covariance structures and acquisition strategies are desired \citep{JonesSchonlauWelch1998,Shahriari2016}.

\SVEMnet can be used with any single-error-stratum experimental design in
which individual runs can be treated as independent. However, the predictive
performance of \SVEMnet will naturally depend on design quality. For physical experiments
we recommend optimal designs
\citep{MyersMontgomeryAndersonCook2016,JonesMontgomery2021DoE}, fast
flexible space-filling designs \citep{LekivetzJones2019FFF}, or U-Bridge
designs \citep{ubridge} rather than the simple Latin hypercube samples
\citep{McKay1979LHS} that are employed for computational expediency in the baseline
simulations in Section~\ref{sec:sim}. Because the
FRW bootstrap treats experimental runs as the resampling units, SVEM is inherently a single-stratum procedure: split-plot and other
multi-stratum designs with hierarchical variance components are not supported and would
require substantial structural changes (e.g., hierarchical base learners and a different
resampling unit). In \SVEMnet{}, blocking or batch indicators may be included only as
fixed additive effects.

The remainder of the paper is organized as follows.
Section~\ref{sec:software} details the \SVEMnet{} modeling engine:
SVEM background, the \glmnet{} wrapper (including relaxed refits),
deterministic expansion tools, the validation-weighted \wAIC/\wBIC
selector, and the multi-response random-search optimizer.
Section~\ref{sec:example} illustrates the workflow on an LNP formulation example.
Section~\ref{sec:sim} reports simulations, and the paper concludes
with a summary and outlook.
Technical specifics of the validation-weighted criterion appear in
\ref{app:criterion}. Supplementary robustness checks and diagnostics appear in
\ref{app:supp}. Mixed-model multiple-comparison tables for simulation results
appear in \ref{app:tukey}.

\section{SVEMnet modeling and multi-response optimization}
\label{sec:software}
\subsection{Software overview}
\label{sec:software-overview}

\SVEMnet takes a user-specified model formula and dataset, generates
fractional random-weight (FRW) training and validation weights, fits elastic-net or
relaxed elastic-net paths with \glmnet{} within each FRW replicate, selects a
penalty (and, when applicable, a relaxed refit) via a validation-weighted
information criterion, and averages predictions across replicates. The same
deterministic expansion of the right-hand side can be reused across responses
so that all models share a common design matrix. The package additionally
provides a permutation-based whole-model test heuristic and a
mixture-constrained random-search optimizer that scores and selects candidates
for additional experimental runs.

The core workflow described here remains available in \SVEMnet{} 3.6.2.
The numerical examples and simulations below retain the original 3.2.0
results; later implementation changes relevant to their interpretation are
identified where needed.

The main functions used here are:
\begin{itemize}
  \item \verb|SVEMnet()|: fit Gaussian or binomial SVEM models under a specified
    formula and dataset, with optional relaxed base learners and FRW control;
  \item \verb|bigexp_terms()| and \verb|bigexp_formula()|: build and reuse
    deterministic high-order factorial and polynomial linear model expansions
    from a single list of main-effects;
  \item \verb|svem_wmt_multi()|: run the permutation-based whole-model test
    (WMT) and obtain approximate $p$-values and importance multipliers for
    multiple responses;
    \item \verb|svem_random_table_multi()|: generate feasible candidate settings
    under mixture and process constraints and compute model predictions;
  \item \verb|svem_score_random()| and \verb|svem_select_from_score_table()|:
    generate feasible candidate settings, score them via multi-response
    desirability, and select diverse optimal and exploration candidates;
  \item \verb|svem_export_candidates_csv()|: export collections of candidate
    tables for laboratory execution or documentation.
\end{itemize}
\subsubsection{Workflow overview and typical use}
\label{sec:workflow}

The within-replicate tuning procedure used to select \((\alpha,\gamma,\lambda)\)
is summarized in Figure~\ref{fig:svem-within-replicate} (Section~\ref{sec:criterion}).
 \SVEMnet{}  embeds this core
engine in a sequential loop: (i) specify the factor space and construct a
deterministic expansion (optional); (ii) fit SVEM models to one or more responses;
(iii) optionally run the permutation-based whole-model diagnostic (WMT;
Section~\ref{sec:wmt}) to assess global signal; (iv) generate a feasible candidate set
under mixture and process constraints; (v) score candidates via multi-response
desirability and summarize high-score (exploitation) and high-uncertainty
(exploration) settings using medoids; and (vi) run selected settings, append the new
rows, and refit. The optional \texttt{wmt\_score} column is treated as a diagnostic
overlay rather than the primary optimization objective.

The remainder of this section explores these functions in more detail. Worked
Gaussian and binomial examples are provided in the CRAN help pages (e.g.,
\verb|?SVEMnet|).

\subsection{SVEM background}
\label{sec:svem-bg}

SVEM draws $B$ replicates using FRW sampling. For each replicate $b$ and
observation $i=1,\dots,n$, training weights $w_i^{\mathrm{train}}$ and an
anti-correlated validation copy $w_i^{\mathrm{valid}}$ are constructed from a
shared $U_i\sim\mathrm{Uniform}(0,1)$ via
\begin{equation}
\label{eq:frw_weights}
w_i^{\mathrm{train}}=-\log U_i,\qquad
w_i^{\mathrm{valid}}=-\log(1-U_i),
\end{equation}
following the FRW scheme for univariate analysis introduced by
\citet{Xu2020_FRW}, which underpins the original SVEM extensions to penalized
regression (and the introduction of the anti-correlated validation weights) by \citet{Gotwalt2018} and \citet{Lemkus2021}. For each FRW
replicate we then rescale both the training and validation weights to have mean
one, $w_i \leftarrow w_i/\bar w$ for
$w\in\{w^{\mathrm{train}},w^{\mathrm{valid}}\}$, matching \glmnet's internal
normalization of observation weights.

Given $(w^{\mathrm{train}},w^{\mathrm{valid}})$ for replicate $b$, \SVEMnet fits an elastic-net or relaxed elastic-net base learner to the FRW-weighted training sample under $w^{\mathrm{train}}$, evaluates a collection of path
points using a validation-weighted information criterion computed with
$w^{\mathrm{valid}}$ (Section~\ref{sec:criterion}), and stores predictions
from the selected path point. Overall SVEM predictions are obtained by
averaging these bootstrap predictions across $b = 1,\dots,B$. \citet{Lemkus2021} tuned penalties by
validation-weighted SSE (\wSSE); \SVEMnet adopts the same
FRW-and-ensemble structure but standardizes on validation-weighted AIC/BIC
analogs---\wAIC default for Gaussian responses; \wBIC default for binomial
responses.

Although we focus on Gaussian responses in the exposition and case study, the
software also implements a binomial (logistic) variant. In that case, the
elastic-net base learner is fit under an FRW-weighted logistic
log-likelihood, penalties are selected on the deviance scale, and ensemble
predictions are formed by averaging bootstrap-member-predicted probabilities; \ref{app:criterion} gives the corresponding criteria.

Related work by \citet{LuAndersonCook2025_NUSFModelSelection} adapts the SVEM
FRW-based self-validation mechanism to define a self-validated mean squared
prediction error (SVMSPE) for model comparison and pairs it with sequential
non-uniform space-filling (NUSF) design augmentation that allocates more runs
to regions where a small set of candidate models differ most in their
predictions. In \SVEMnet{}, the FRW validation losses instead feed
validation-weighted AIC/BIC-type selectors within a given elastic-net SVEM
model class, while sequential design is handled separately by the
desirability-based random-search optimizer in Section~\ref{sec:optimizer}.

\subsection{Implementation note: a SVEM wrapper over \glmnet}

\SVEMnet implements SVEM by wrapping the standard and relaxed elastic-net
paths in \glmnet{} \citep{en05,ESL2009,Friedman2010_glmnet,Friedman2025_relax}
with FRW train/validation weights (Section~\ref{sec:svem-bg}). For each FRW
replicate $b$ and each $\alpha$ in a user-specified grid (default
$\alpha\in\{0.5,1\}$), \SVEMnet calls \glmnet{} with training weights
$w^{\mathrm{train}}$, an intercept, and predictor standardization enabled. By default, \SVEMnet uses relaxed paths for Gaussian responses and
standard elastic-net paths for binomial responses.

Relaxation alters the usual bias--variance trade-off by reducing shrinkage after
variable selection: individual relaxed fits can have lower bias but higher variance,
especially near the interpolation boundary. In a SVEM ensemble, averaging across FRW
replicates can partially stabilize this additional variability. Empirically, the
Gaussian simulations in Section~\ref{sec:sim} favor relaxed SVEM fits, whereas the
binomial simulations favor non-relaxed fits; we default to
\texttt{relaxed = TRUE} for Gaussian and \texttt{relaxed = FALSE} for binomial
responses. \ref{app:supp_relax} provides additional diagnostic summaries of
selected $(\lambda,\gamma)$ distributions and bias--variance behavior in representative
Gaussian and binomial settings.

Within each replicate, \SVEMnet treats each combination of $\alpha$ and
penalty $\lambda$ (and, when relaxed fits are enabled, each relaxed refit value
$\gamma$) as a candidate path point. For a given \glmnet{} fit, the
$\lambda$ path is traversed and, for relaxed fits, a small grid of $\gamma$
values is considered. Each candidate $(\alpha,\lambda,\gamma)$ is evaluated on
the FRW validation copy using the chosen validation-weighted criterion
(\wAIC, \wBIC, or \wSSE; \ref{app:criterion}), and the combination
that minimizes this criterion determines the selected coefficients for
replicate $b$. These selected coefficients populate a
bootstrap-by-parameter coefficient matrix.

For new data, the design matrix is rebuilt to match training, each replicate's
coefficients are applied, and the resulting bootstrap predictions are
averaged---on the response scale for Gaussian models and, by default, on the
probability scale for binomial models. For Gaussian models, an optional \emph{debias} step fits a linear calibration
$\mathrm{lm}(y \sim \hat y)$ on the training data and applies the resulting
intercept and slope to ensemble predictions. Debiasing is off by default and
is included mainly to match the JMP implementation, where it is always
applied \citep{JMP19SVEM}.

\subsubsection{Deterministic expansion utilities}

A practical convenience in \SVEMnet is that the user lists main-effect
predictors only once and lets the expansion utilities (\lstinline|bigexp_terms()|
and \lstinline|bigexp_formula()|) deterministically build interaction and
polynomial columns. The same specification can be reused across responses,
guaranteeing identical design matrices, stable column order, and reproducible
comparisons. With \texttt{factorial\_order = 2} and
\texttt{polynomial\_order = 2}, for example, the generated right-hand side
includes all main effects, two-way interactions, and quadratic terms (and, optionally,
partial-cubic terms of the form $\mathrm{I}(X^2)\!:\!Z$) without hand-writing
products or powers; categorical factors are handled using the contrasts
recorded for the training data.  We also include any mixture factors in the full expansion rather than constructing Scheff\'e polynomials \citep{scheffe}.

When blocking or covariate factors (e.g., Operator, day, plate, ambient temperature) are
supplied via \texttt{blocking} in \texttt{bigexp\_terms()}, these factors make an additive adjustment to the response surface and do not interact with the other study factors. 

\paragraph{Discrete numeric predictors}

Some process variables are recorded as numeric but are only feasible to run at a discrete set
of levels (e.g., fixed pump settings or allowable N:P ratios). To avoid proposing
unattainable intermediate settings in candidate generation and whole-model diagnostics,
\SVEMnet{} allows users to mark such predictors via the \texttt{discrete\_numeric}
argument in \lstinline|bigexp_terms()| (which affects candidate and WMT evaluation settings, not the SVEM model fit). The
resulting specification records the allowed numeric support in the training data, and
downstream tools (random candidate generation and WMT evaluation-set construction)
restrict sampling to these observed levels.

\paragraph{Practical guidance on choosing expansion order}

When the true response surface order is unknown, we recommend starting with a
moderately rich expansion that is scientifically plausible (e.g., third-order
polynomials with up to three-way interactions and partial cubic terms for mixture--process screening), and
allowing the elastic-net SVEM fit to downselect.  This ``start wide, shrink down''
strategy is consistent with the SVEM-based LNP workflow of \citet{JoVE_LNP} and is
supported by the simulation results in Section~\ref{sec:sim}, where mild
over-specification is typically less harmful than gross under-specification.
In practice we recommend: (i) fit a candidate expansion; (ii) examine
actual-by-predicted diagnostics and predictive uncertainty; (iii) optionally apply WMT
(Section~\ref{sec:wmt}) to assess global signal; and (iv) if results are unstable,
compare two adjacent expansion orders (e.g., Orders~2 and~3) and check whether the top
candidate sets and fitted predictions are materially changed. A simple candidate
stability check is to re-run scoring under both expansions and report the overlap of
top-ranked recipes and medoids (\ref{app:supp_modelorder}).

\subsection{Selection criterion}
\label{sec:criterion}

\begin{figure}[!t]
  \centering
  \includegraphics[width=1\linewidth]{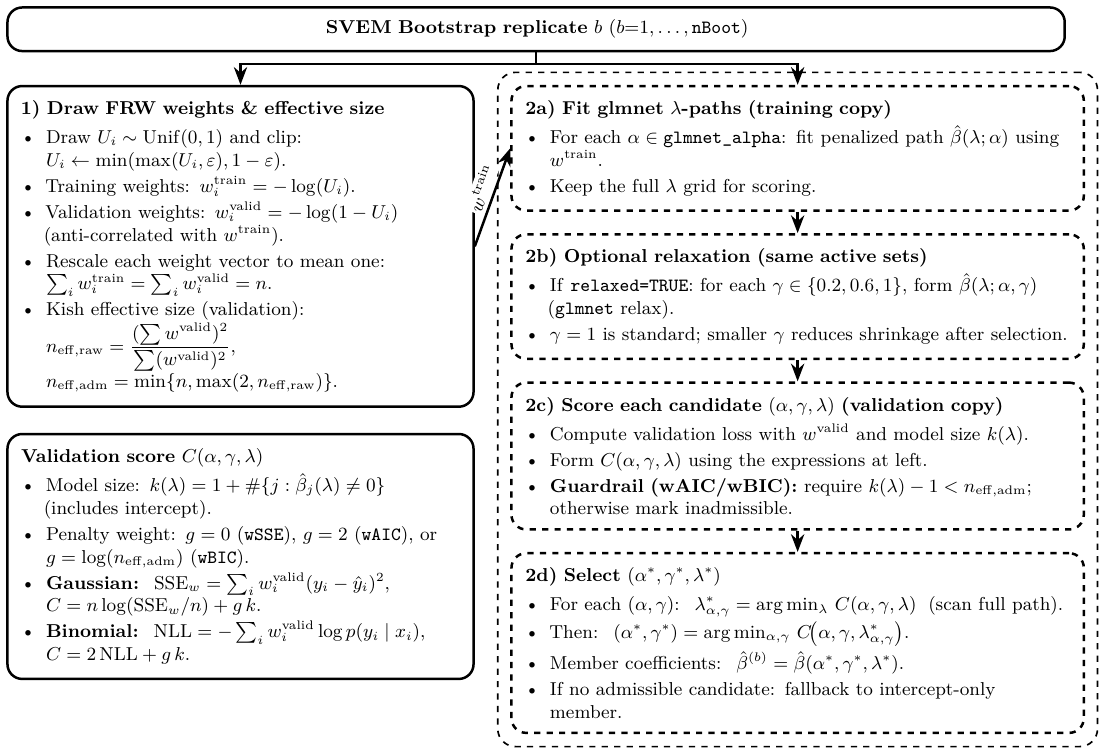}
  \caption{Within-replicate SVEMnet tuning flow for one bootstrap member \(b\).}
  \label{fig:svem-within-replicate}
\end{figure}

At each candidate path point within an FRW replicate---indexed by the
elastic-net mixing parameter $\alpha$, the penalty $\lambda$, and, when
relaxed fits are used, a relaxation parameter $\gamma$---\SVEMnet evaluates a
validation loss on the anti-correlated FRW copy. For Gaussian responses,
the information-criterion loss is $n\log\{\mathrm{SSE}_w/n\}$; for binomial
responses it is twice the weighted negative log-likelihood. We add a penalty
$g\,k_\lambda$, where $k_\lambda$ counts the intercept plus the nonzero slope
coefficients. Without this penalty, minimizing the Gaussian log-loss is
equivalent to minimizing the weighted sum of squared errors (\wSSE).
Choosing $g = 2$ or $g = \log(n_{\mathrm{eff}}^{\mathrm{adm}})$ gives
validation-weighted analogs of AIC (\wAIC) and BIC (\wBIC), respectively. \ref{app:criterion} records the full expressions and defines the
effective-size guardrail $n_{\mathrm{eff}}^{\mathrm{adm}}$ used with \wAIC{}
and \wBIC{} to avoid near-interpolating path points. This admissibility guardrail is applied only for the information-criterion
selectors (\wAIC{}, \wBIC{}); the loss-only \wSSE{} selector evaluates all path
points without this constraint.  Figure~\ref{fig:svem-within-replicate} summarizes the algorithm.

Because the FRW validation weights are random rather than fixed design
weights, these scores are used heuristically for relative model comparison
inside the FRW replicate rather than as exact AIC/BIC. As noted by
\citet{LuAndersonCook2025_NUSFModelSelection}, classical uses of AIC and BIC
evaluate fit at the same observed locations used for estimation, which can
lead to over-optimistic assessments. In \SVEMnet{}, the \wAIC{} and \wBIC{}
criteria are always computed on the FRW validation copy. For binomial
models we retain the label \wSSE{}  even though the
selector operates on the deviance scale rather than squared error.

\paragraph{Heuristic status and boundary behavior}

Because the validation copy is itself random under FRW resampling, \wAIC{} and
\wBIC{} are best interpreted as validation-weighted tuning scores rather than as
classical information criteria with formal asymptotic guarantees. When
$w_i^{\mathrm{valid}}\equiv 1$ these criteria reduce (up to constants independent of
$\lambda$) to the standard Gaussian AIC/BIC forms, and the use of
$n_{\mathrm{eff}}^{\mathrm{adm}}$ in the BIC-style penalty can be viewed as a simple
effective-size adjustment for highly variable validation weights. Empirically, the
primary motivation for \wAIC{} and \wBIC{} in SVEM is stability near the
interpolation boundary $n\approx\pfull$. \ref{app:supp_boundary} provides a
focused simulation around $n=\pfull$ that compares \wAIC/\wBIC with the loss-only
\wSSE{} selector in terms of selected model size and out-of-sample error.

\subsection{Permutation-based whole-model diagnostic (WMT)}
\label{sec:wmt}

\SVEMnet{} includes a permutation-based whole-model test (WMT) heuristic for Gaussian
SVEM fits \citep{Karl2024_svem_test}. The goal is to provide a practical diagnostic
for whether a fitted SVEM model exhibits global factor signal beyond what would be
expected under a constant-response null, complementing the usual actual-by-predicted
and residual plots. WMT is a randomized, simulation-based procedure and the resulting
``$p$-values'' are approximate; we treat them as diagnostic summaries rather than as
formal hypothesis tests.

At a high level, WMT compares standardized SVEM predictions on a space-filling set of
evaluation points with a reference distribution obtained by refitting the same SVEM
specification to permuted responses. Let $T=\{x_i\}_{i=1}^{n_{\mathrm{point}}}$ be a
space-filling set over the modeled factor space (respecting mixture constraints,
categorical levels, and any recorded discrete numeric supports). For each evaluation
point $x\in T$, we compute a standardized prediction
\begin{equation}
\label{eq:wmt_standardized}
h(x) = \frac{\hat f(x) - \bar y}{\hat s(x)},
\end{equation}
where $\hat f(x)$ is the SVEM ensemble mean prediction, $\bar y$ is the training-set
mean response, and $\hat s(x)$ is the ensemble standard deviation across SVEM bootstrap
predictions at $x$ (a pointwise uncertainty proxy) \citep{Karl2024_svem_test}. Repeating this calculation for
$n_{\mathrm{SVEM}}$ independent SVEM refits (new FRW weights each time) yields an
$n_{\mathrm{SVEM}}\times n_{\mathrm{point}}$ matrix of standardized predictions.
WMT then fits the same SVEM specification to $n_{\mathrm{perm}}$ random permutations of
the response vector and records the corresponding standardized predictions on $T$ in an
$n_{\mathrm{perm}}\times n_{\mathrm{point}}$ reference matrix.

To summarize global deviation from the null, WMT standardizes the permutation matrix
columnwise, applies a reduced-rank singular value decomposition (retaining the leading
components explaining a user-specified fraction of the permutation variance), and
computes Mahalanobis-like distances for the permutation rows and for the original-data
refits, following \citet{Karl2024_svem_test}. The resulting distance distributions
are displayed graphically; clear separation indicates strong global signal. For a
scalar summary, \SVEMnet{} fits a flexible parametric distribution to the permutation
distances (SHASH \citep{shash}) and reports the fitted right-tail probability as a  whole-model ``$p$-value.'' Because the procedure uses finite budgets
($n_{\mathrm{perm}}$, $n_{\mathrm{point}}$, $n_{\mathrm{SVEM}}$), results have Monte
Carlo variability; for interpretability, users should set and report random seeds and
treat marginal cases cautiously. The current \SVEMnet{} defaults are
$n_{\mathrm{point}}=2000$, $n_{\mathrm{SVEM}}=10$, $n_{\mathrm{perm}}=150$, and a
90\% variance-retention threshold for the reduced-rank distance calculation.
The WMT has separate inner-fit controls: by default each original or permuted
fit uses 100 bootstrap members, $\alpha=1$, and no relaxation. It does not
automatically reuse the fitted predictive ensemble.

In \SVEMnet{}, WMT can be run for multiple responses via
\lstinline|svem_wmt_multi()|, which reuses a shared deterministic expansion and
constraint specification. The resulting WMT ``multipliers'' (monotone transforms of
the per-response approximate $p$-values) can optionally reweight user-specified
multi-response desirability weights when producing the diagnostic \texttt{wmt\_score}
column (Section~\ref{sec:optimizer}). In this paper we treat WMT reweighting as a
secondary sensitivity check rather than as the primary decision rule, consistent with
the diagnostic intent of the test heuristic.

\subsection{Random-search scoring and candidate selection for multi-response design}
\label{sec:optimizer}

The built-in stochastic optimizer implements the desirability-based
generate-and-rank strategy sketched in the introduction. Starting from fitted
\SVEMnet{} models, it generates feasible candidate settings under the observed
factor ranges and user-specified mixture constraints, evaluates each setting
via a multi-response desirability score, and returns shortlists of
high-score (exploitation) and high-uncertainty (exploration) candidates. The
following subsections describe candidate generation and mean-level
specification probabilities, construction and weighting of desirability
scores, the uncertainty measure used to target exploration, and the
medoid-based selection of diverse recipes.

\subsubsection{Sampling and specification limits}

Candidate settings and their model predictions are generated by
\texttt{svem\_random\_table\_multi()}. In typical use,
\texttt{svem\_score\_random()} calls this generator and appends desirability
scores, uncertainty summaries, and optional mean-in-specification proxies.
For each response with supplied limits, the package records the proportion of
bootstrap-member mean predictions inside those limits in a column such as
\texttt{Potency\_p\_in\_spec\_mean}. The column \texttt{p\_joint\_mean}
is the product of these response-wise proportions. It is a ranking proxy,
not a calibrated joint probability accounting for dependence between responses.
These summaries concern modeled means and do not estimate conformance rates
for individual future observations.

\subsubsection{Scoring and response weighting}

For each response $r$, we map predictions to unit-interval desirabilities
$d_r(x)\in[0,1]$ using Derringer--Suich curves 
and form a multi-response score as a weighted geometric mean of the
per-response desirabilities,
\begin{equation}
\label{eq:desirability_score}
S(x)
= \exp\!\left(\sum_r w_r \log\big(d_{r,\varepsilon}(x)\big)\right), \qquad
d_{r,\varepsilon}(x) = (1-\varepsilon)\,d_r(x) + \varepsilon.
\end{equation}
Here a small $\varepsilon$ (default $10^{-6}$) avoids $\log(0)$ \citep{Harrington1965,DerringerSuich1980,MyersMontgomeryAndersonCook2016}.

\paragraph{Practical desirability calibration and sensitivity checks}

In practice we recommend choosing Derringer--Suich parameters from scientific
acceptability ranges rather than from the fitted model itself. For each response, the
user specifies a goal (maximize, minimize, or hit a target) together with bounds and
optionally a target. The bounds define where desirability is essentially zero (clearly
unacceptable) and where it saturates at one (fully acceptable or ``good enough''); the
shape exponent controls how rapidly desirability approaches one between these values.
\SVEMnet{} uses a simple default shape (exponent $=1$) unless the user supplies
custom exponents, which yields a linear ramp on the response scale. Exponents greater
than one make the desirability curve more stringent near the target, while exponents
below one make it more tolerant.

The weighted geometric mean in~\eqref{eq:desirability_score} is used because it is the
standard aggregation in the desirability literature and penalizes low
desirability more strongly than an arithmetic mean \citep{DerringerSuich1980}.
With the positive floor in~\eqref{eq:desirability_score}, however, a zero
desirability contributes a factor $\varepsilon^{w_r}$ rather than forcing
the overall score to zero. Thus the score ranks trade-offs; hard acceptability
requirements should be enforced separately.

When desirability parameters encode subjective trade-offs, simple sensitivity checks are
valuable. A practical approach is to rerun scoring under plausible alternative bounds,
targets, exponents, or response weights and then compare the overlap and rank stability
of the top candidates and medoids. \ref{app:supp_desirability} provides a
compact desirability-calibration sensitivity check.

User-specified goal weights $\tilde w_r$ are first normalized to
$\bar w_r$ with $\sum_r \bar w_r = 1$. When
\texttt{wmt = NULL}, we set $w_r = \bar w_r$ and report a
single column \texttt{score} based on $S(x)$. When a WMT result is supplied
through \texttt{wmt} for Gaussian responses, we further
reweight the normalized user weights by a monotone function of the
permutation whole-model $p$-values from \citep{Karl2024_svem_test}
(for example $-\log_{10}(p_r)$), renormalize to obtain WMT-adjusted weights
$w_r^{\mathrm{WMT}}$, and compute a second geometric-mean score
\texttt{wmt\_score} by using $w_r = w_r^{\mathrm{WMT}}$ in $S(x)$.
These WMT-adjusted weights are diagnostic and do not
alter the underlying SVEM fits. In this paper we treat the resulting WMT-based
multipliers as a secondary priority: the primary optimization uses
the user-specified weights $\tilde w_r$, and \texttt{wmt\_score} is intended
to provide additional, lower-priority optimal candidates when budget allows.

Thus the output table always contains \texttt{score}, based purely on the
user-specified weights $\tilde w_r$, and, when applicable, an additional
\texttt{wmt\_score} column that augments those weights using the WMT results. Gaussian and binomial outcomes are handled on the same
scale by applying desirability functions to predicted means and predicted
probabilities, respectively.

For candidate scoring, \texttt{blocking} variables are held fixed at a single reference
setting while the controllable factors are varied over the feasible region.
Categorical blocking factors are evaluated at the most common observed level, and
continuous blocking covariates are evaluated at the midpoint of their modeled range.
Under the additive-only blocking assumption this defines a consistent reference for
comparing controllable settings; if desirability bounds/targets are specified on an
absolute response scale, users may optionally re-score at alternative blocking levels
to confirm stability of the top candidates.

\subsubsection{Uncertainty and exploration}

For each response $r$ we compute a central percentile interval
$[\ell_r(x),u_r(x)]$ of the bootstrap-member predictions at the requested
\texttt{level} (default 0.95) and define its width
$W_r(x) = u_r(x) - \ell_r(x)$. These intervals
summarize variation among members fitted to the same data, which need not
represent the sampling uncertainty of the complete ensemble predictor. They
neither establish confidence coverage for the true mean response nor include
the future-observation noise required for prediction intervals. To compare
responses, we normalize $W_r(x)$ using the empirical 2\% and 98\% quantiles
of interval widths across the sampled candidates and aggregate using the
original normalized goal weights $\bar w_r$:
\begin{equation}
\label{eq:uncertainty_measure}
U(x)
= \sum_r \bar w_r\, \mathrm{Norm}_{0,1}\!\big(W_r(x);\; q_{r,0.02}, q_{r,0.98}\big),
\end{equation}
where $\mathrm{Norm}_{0,1}(\cdot; q_{r,0.02}, q_{r,0.98})$ denotes a
linear rescaling clipped to $[0,1]$, using the width quantiles
$q_{r,0.02}$ and $q_{r,0.98}$ for response $r$. Large values of $U(x)$
identify candidates with greater member variability relative to the sampled
set. The selection function returns one maximizing row (ties can occur)
and can also use $U(x)$ to construct the exploration shortlists
described in the next subsection.

\subsubsection{Diverse candidates via medoids and practical defaults}

\begin{figure}[!t]
  \centering
  \includegraphics[width=.70\textwidth]{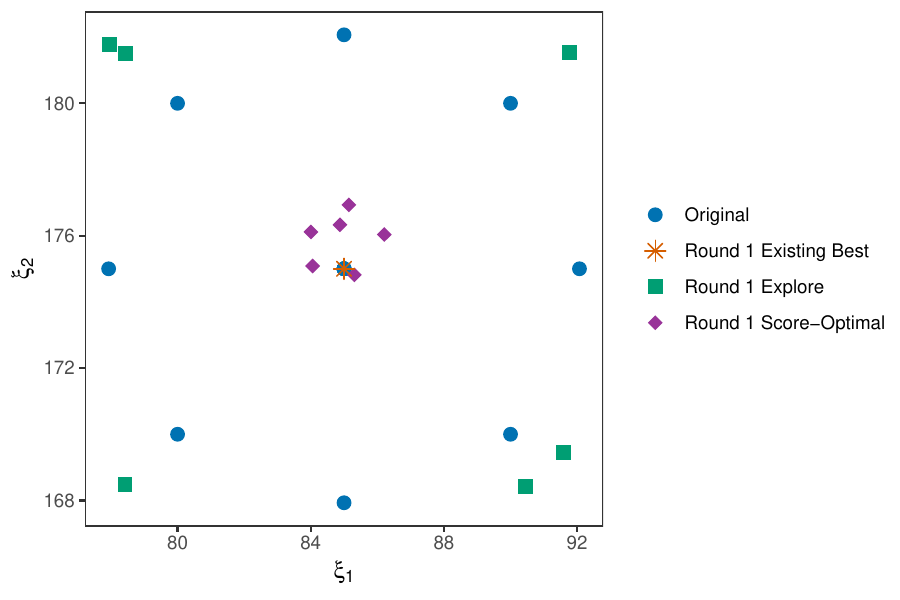}
  \caption{Original CCD runs from Table~6.8 of
    \citet{MyersMontgomeryAndersonCook2016}
    (blue circles), SVEMnet-selected optimal candidates (purple diamonds),
    and high-uncertainty exploration runs (green squares) for the $y_1$
    response, which attains its maximum near the center of the design region.}
  \label{fig:ccd}
\end{figure}

Score and uncertainty surfaces often exhibit ridges of near-equivalent values
across a broad range of factor settings. To capture this range of feasible
recipes with similar predicted performance, the user may choose a number $k$
of candidates by ranking the scored table on a scalar objective (typically
$S(x)$ or $U(x)$, corresponding to the \texttt{score} and
\texttt{uncertainty\_measure} columns, but also, for example,
\texttt{wmt\_score}, \texttt{p\_joint\_mean}, or any other numeric column),
retaining either the top fraction or the top $n$ rows, and clustering this
subset using Gower distance on the predictor columns with
partitioning-around-medoids (PAM) \citep{Gower1971,KaufmanRousseeuw1990}. The
medoids are existing feasible rows and summarize diverse ``exploitation'' and
``exploration'' recipes. The selection function
\texttt{svem\_select\_from\_score\_table()} also returns a best row under
the chosen objective. Figure~\ref{fig:ccd}
illustrates this strategy for the central composite design from Table~6.8 of
\citet{MyersMontgomeryAndersonCook2016}.

\section{Example LNP Application}\label{sec:example}

We illustrate the workflow using the bundled simulated LNP formulation screening data,
which include four mixture components (PEG, Helper, Ionizable, Cholesterol), a
categorical ionizable lipid type, two discrete numeric process factors (\texttt{N\_P\_ratio} and
\texttt{flow\_rate}), a categorical Operator factor, and three responses: Potency, Size, and PDI. This example
parallels the SVEM-based mixture--process LNP workflow of \citet{JoVE_LNP}.

We first build a single
deterministic right-hand-side expansion and reuse it across all three models. We treat Operator as a blocking variable that enters additively and does not interact with other study factors.

\begin{lstlisting}
library(SVEMnet)
data(lipid_screen)
spec <- bigexp_terms(
  Potency ~ PEG + Helper + Ionizable + Cholesterol +
    Ionizable_Lipid_Type + N_P_ratio + flow_rate,
  blocking         = "Operator",  # additive blocking factor  
  discrete_numeric = c("N_P_ratio","flow_rate"),  
  data             = lipid_screen,
  factorial_order  = 3,   # up to 3-way interactions
  polynomial_order = 3,   # up to cubic terms 
  include_pc_2way  = TRUE
)
form_pot <- bigexp_formula(spec, "Potency")
form_siz <- bigexp_formula(spec, "Size")
form_pdi <- bigexp_formula(spec, "PDI")
\end{lstlisting}

Using this shared expansion for a third-order model, we fit Gaussian SVEM models for each response with
default settings and collect them in a named list. Multi-response goals are set
so that Potency is maximized, while Size and PDI are minimized, and mixture
constraints reflect practical bounds and a unit-sum constraint on the four
composition variables. This example uses the default desirability anchors
derived from the candidate predictions; the separate specification limits
introduced below do not set those anchors. Scientific acceptability bounds
can instead be supplied explicitly in \texttt{goals}.

\begin{lstlisting}
set.seed(1)
fit_pot <- SVEMnet(form_pot, lipid_screen)
fit_siz <- SVEMnet(form_siz, lipid_screen)
fit_pdi <- SVEMnet(form_pdi, lipid_screen)

objs <- list(Potency = fit_pot, Size = fit_siz, PDI = fit_pdi)

goals <- list(
  Potency = list(goal = "max", weight = 0.6),
  Size    = list(goal = "min", weight = 0.3),
  PDI     = list(goal = "min", weight = 0.1)
)

mix <- list(list(
  vars  = c("PEG", "Helper", "Ionizable", "Cholesterol"),
  lower = c(0.01, 0.10, 0.10, 0.10),
  upper = c(0.05, 0.60, 0.60, 0.60),
  total = 1.0
))
\end{lstlisting}

\subsection{Whole-model reweighting and permutation tests}

The numerical WMT outputs in this example and \ref{app:supp_wmt}
are the archived 3.2.0 results. Later releases changed the treatment of
blocking variables in the evaluation grid, the sampling RNG, and the numerical
SHASH fit. The historical values are therefore not current-release
reproduction targets and should not be interpreted as block-adjusted tests
of the controllable factors alone.

To assess which responses show the strongest relationship with the study factors, we apply the
permutation-based whole-model test heuristic (WMT) of \citet{Karl2024_svem_test} to each
SVEM model. WMT-based multipliers can optionally reweight the user-specified
response weights, prioritizing responses with stronger predictive signal and
down-weighting weaker ones. Consistent with \citet{Karl2024_svem_test}, we
view these results as a way to highlight responses with clearer global signal,
not as a hard filter: in small or noisy studies the WMT heuristic can have
limited power, so candidate sets should not be chosen solely by appealing to
WMT $p$-values or multipliers. Because WMT is permutation-based with finite
simulation budgets, these $p$-values are approximate.

The helper \lstinline|svem_wmt_multi()| runs the permutation test for each
response under the shared deterministic expansion and mixture constraints, and
returns approximate $p$-values and multipliers:

\begin{lstlisting}
set.seed(123)
wmt_out <- svem_wmt_multi(
  formulas       = list(Potency = form_pot,
                        Size    = form_siz,
                        PDI     = form_pdi),
  data           = lipid_screen,
  mixture_groups = mix,
  wmt_control    = list(seed = 123)
)
wmt_out$p_values
wmt_out$multipliers
\end{lstlisting}

\begin{figure}[!t]
  \centering
  \includegraphics[width=.60\textwidth]{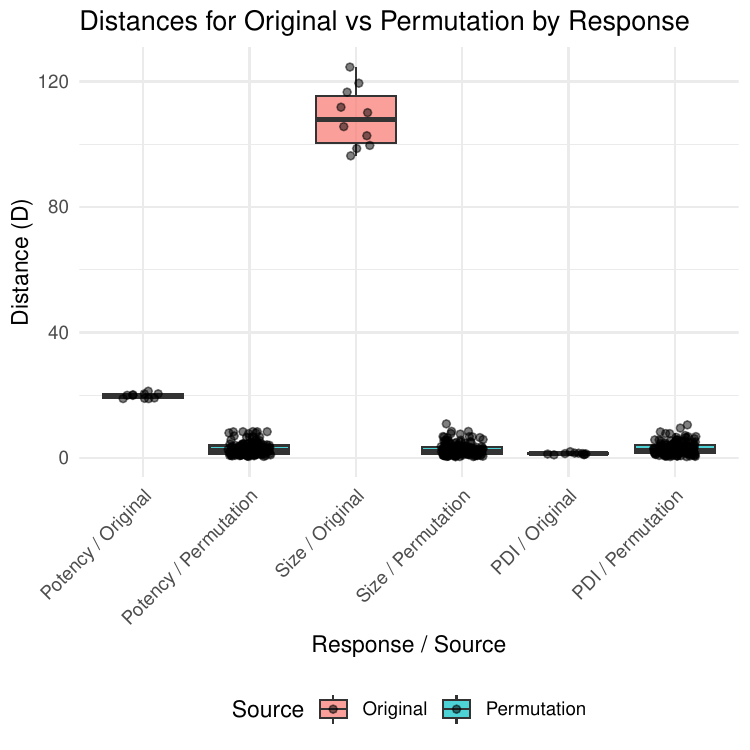}
  \caption[Whole-model significance diagnostic in \SVEMnet]{Whole-model
  significance diagnostic (WMT) for Potency, Size, and PDI: distance
  distributions for the original data versus permuted fits under the shared
  expansion and mixture constraints.}
  \label{fig:wmt_karl_style}
\end{figure}

The approximate whole-model $p$-values under the default permutation budget were
Size $p\approx 1.0\times 10^{-16}$, Potency $p\approx 4.1\times 10^{-9}$, and
PDI $p\approx 0.77$. Thus Size and Potency show strong global signal, whereas PDI does
not (although lack of signal in this heuristic alone should not be used to conclude
that there is no relationship between the study factors and PDI
\citep{Karl2024_svem_test}). Figure~\ref{fig:wmt_karl_style} shows the associated
whole-model significance diagnostic, generated by \lstinline|svem_wmt_multi()| under
the shared expansion and mixture constraints. Clear separation between the
``Original'' and permutation-based distance distributions indicates strong global
factor signal, as in Figure~3 of \citet{Karl2024_svem_test}. \ref{app:supp_wmt}
reports a simple stability check over random seeds and permutation budgets for this
example.

\subsection{Random-search scoring, candidate selection, and sequential use}

To illustrate the optional specification-limit handling, we impose mean-level
requirements on the three responses---limits on the predicted process means
rather than on individual unit-level measurements: Potency $>78$, Size $<100$, and PDI $<0.25$. Only the bounded side of each limit
needs to be supplied:

\begin{lstlisting}
specs_ds <- list(
  Potency = list(lower = 78),
  Size    = list(upper = 100),
  PDI     = list(upper = 0.25)
)
\end{lstlisting}

The random-search scoring step then evaluates a large number of feasible
settings ($25{,}000$ in this example). For
each candidate, \lstinline|svem_score_random()| evaluates the SVEM models,
computes per-response Derringer--Suich desirabilities, aggregates them into a
single multi-response score via a weighted geometric mean, constructs a scalar
uncertainty measure $U(x)$ from bootstrap percentile intervals, and, when specifications are supplied, computes mean-level probabilities that
each candidate's modeled mean lies inside the user limits.
Supplying \lstinline|wmt = wmt_out| adds a WMT-adjusted score column
\lstinline|wmt_score| in addition to the user-weighted \lstinline|score|:

\begin{lstlisting}
set.seed(3)
scored <- svem_score_random(
  n              = 25000,
  objects        = objs,
  goals          = goals,
  data           = lipid_screen,
  mixture_groups = mix,
  wmt            = wmt_out,   # optional: NULL for no WMT
  specs          = specs_ds   # optional: NULL for no design-space columns
)
\end{lstlisting}
From the resulting scored table we extract high-score, exploration, and
design-space candidates using \lstinline|svem_select_from_score_table()|.
For example, the following call selects five near-optimal score medoids, in addition to the single highest scored recipe from the candidate set of $25{,}000$:

\begin{lstlisting}
# Score-optimal medoids (user-weighted score)
opt_sel <- svem_select_from_score_table(
  score_table = scored$score_table,
  target      = "score",
  direction   = "max",
  k           = 5,
  top_type    = "frac",
  top         = 0.1,
  label       = "round1_score_optimal"
)
\end{lstlisting}
Analogous calls targeting \texttt{uncertainty\_measure} or
\texttt{p\_joint\_mean}, applied to the random-search candidates or the
original screening runs, produced the candidates in
Tables~\ref{tab:lnp-cands-recipe} and~\ref{tab:lnp-cands-behav}.
The replication script contains the full set of \texttt{svem\_select\_from\_score\_table()} calls \citep{Karl2025_SVEMnetData}. Candidate tables for laboratory use can be exported
with \lstinline|svem_export_candidates_csv()|; the Supplementary CSV file was
generated using this function and is included in the Mendeley Data archive
\citep{Karl2025_SVEMnetData}.

Tables~\ref{tab:lnp-cands-recipe} and~\ref{tab:lnp-cands-behav} show four
SVEM candidates from the simulated first-round lipid screening. The first
row corresponds to the best existing screened run by multi-response score and
provides a baseline. This row displays the original level of the \texttt{blocking} factor, Operator, which may differ from the level shown for the random-search candidates; the latter are evaluated at the reference Operator level used for scoring. The additive offsets for the Operator levels are visible in the fitted coefficients (e.g., \verb|coef(fit_pot)|), and predictions for other Operator levels can be obtained via \verb|predict(fit_pot, newdata = ...)|. The score-optimal run from the random
search trades a modest reduction in predicted Potency for a large improvement
in Size, yielding an overall desirability close to one. In this example, the
\texttt{wmt\_score} optimum coincides with the score-optimal run, so no additional
row is shown. The ``In-spec'' optimum illustrates a candidate that lies
comfortably inside the specification region. The exploration target
has the highest uncertainty measure $U(x)$ and sits near the edge of the
feasible mixture region, with poor predicted Potency. Table~\ref{tab:lnp-cands-behav} reports the SVEM
point predictions together with the primary desirability score (\texttt{score}), the
WMT-reweighted diagnostic score (\texttt{wmt\_score}), the mean-in-spec proxy
$p_{\mathrm{joint,mean}}$, and the uncertainty measure~$U(x)$
(Section~\ref{sec:optimizer}).

\begin{table}[!htbp] 
\centering
\small
\caption{Representative SVEM candidate recipes from the lipid screening
example. Mixture components are proportions on the 0--1 scale; process
factors are in their native units. All random-search candidates are evaluated at the reference (modal) Operator level A.}
\label{tab:lnp-cands-recipe}
\begin{tabular}{lccccccccc}
\toprule
Scenario & PEG & Helper & Ionizable & Cholesterol &
Ionizable Type & N:P & flow rate & Operator\\
\midrule
Best screened  & 0.047 & 0.430 & 0.140 & 0.383 & H101 &  6 & 2.5&B \\
Max Score      & 0.048 & 0.583 & 0.110 & 0.259 & H101 & 10 & 2.5&A \\
In-spec            & 0.040 & 0.458 & 0.165 & 0.337 & H103 & 14 & 2.5&A \\
Exploration & 0.011 & 0.592 & 0.123 & 0.273 & H103 & 6 & 2.5&A \\
\bottomrule
\end{tabular}
\end{table}

\begin{table}[!htbp]
\centering
\small
\caption{Predicted behavior for the candidate recipes in
Table~\ref{tab:lnp-cands-recipe}. Responses are SVEM point predictions.}
\label{tab:lnp-cands-behav}
\begin{tabular}{lcccccccc}
\toprule
Scenario & Potency & Size & PDI & score & wmt\_score &
$p_{\mathrm{joint,mean}}$ &  $U(x)$ \\
\midrule
Best screened    & 91.5 &  88.2 & 0.21 & 0.83 & 0.88 & 0.93 & 0.77 \\
Max Score        & 89.1 &  50.4 & 0.17 & 1.00 & 1.00 & 1.00 & 0.39 \\
In-spec          & 84.9 &  76.5 & 0.17 & 0.83 & 0.82 & 1.00 & 0.73 \\
Exploration      & 74.3 & 51.2 & 0.20 & 0.45 & 0.54 & 0.09 & 0.94 \\
\bottomrule
\end{tabular}
\end{table}

\subsection{Computational environment and runtime}

All timings were obtained on a Windows 11 x64 desktop with an
Intel Core Ultra 7 265 CPU
(20 cores, 2.4~GHz) and 64~GB RAM, running R 4.5.2 with \SVEMnet{} 3.2.0 and
\glmnet{} 4.1-10 and parallel execution enabled for the WMT. Users should note that on Windows, parallelization of the WMT may trigger a one-time Windows Firewall
prompt for R when worker processes are created; this is due to local socket connections between R
processes and does not involve external network access. In typical setups the computation will still
run if the prompt is dismissed.

 For the lipid screening
example (23 experimental runs, three Gaussian responses, shared third-order
deterministic expansion, and a 25\,000-point random candidate set), a single
end-to-end run of the workflow required about 260~seconds: roughly 92~seconds
to fit \SVEMnet for the three responses, 156~seconds for the
optional permutation-based whole-model tests, and 9~seconds for the
random-search scoring, candidate selection, and CSV export. Thus most of the
computational cost lies in SVEM model fitting and the WMT; the optimization
layer is comparatively light.

\section{Simulation study}\label{sec:sim}

We carried out simulations comparing \SVEMnet to repeated
cross-validated elastic-net baselines over grids of sample sizes and signal
strengths, where ``signal strength'' refers to the proportion of variance in the
data explained by the underlying model (target $R^2$). The baselines use a wrapper \texttt{glmnet\_with\_cv()} around \texttt{cv.glmnet}, matching the
$\alpha$ grid and penalty path to the \SVEMnet settings. For each configuration
we ran both the default two-point grid $\alpha\in\{0.5,1\}$ and a lasso-only grid
$\alpha=1$, which behaved indistinguishably in these simulations. All simulation scripts and CSV result files are provided in the replication bundle of
\citet{Karl2025_SVEMnetData}.

\paragraph{Version scope of the repeated-CV comparisons}
The reported simulations use \SVEMnet{} 3.2.0. Its relaxed
\texttt{glmnet\_with\_cv()} wrapper selected $\lambda$ from the
$\gamma=1$ CV curves and combined it with a $\gamma$ selected in a separate
final CV fit. Version 3.4.0 corrected this procedure to aggregate the full
relaxed CV surfaces and jointly select $(\alpha,\lambda,\gamma)$.
Consequently, the relaxed-CV rows and relaxation diagnostics reported here
describe the historical wrapper, not the corrected joint selector. They do
not establish that relaxation itself worsens properly tuned repeated CV.
The non-relaxed CV curves in the overview plots are outside the scope of
this relaxed-selector correction, as is the SVEM relaxation search.

\subsection{Design and metrics}

Each replicate uses a sparse quadratic response surface in four continuous factors
$X_1{:}X_4 \in [-1,1]$ and one three-level factor
$X_5 \in \{\mathrm{L1},\mathrm{L2},\mathrm{L3}\}$ (sum-to-zero coding). Training data are Latin hypercube samples for $X_1{:}X_4$ with balanced $X_5$ levels \citep{McKay1979LHS}, with
$n_{\text{total}} \in \{15,20,\dots,50\}$.  As noted in the introduction, we use Latin hypercubes here for simulation convenience; in practice, more sophisticated optimal or space-filling designs are typically preferable when constructing the initial experimental design \citep{Cornell2002,MyersMontgomeryAndersonCook2016,LekivetzJones2019FFF,ubridge,JonesMontgomery2021DoE}.  
\ref{app:supp_ccd} reports representative
robustness checks using a face-centered central composite design (CCD) in place of the
Latin-hypercube baseline and using denser/alternative coefficient regimes.
All simulations treat runs as independent (single error stratum); hierarchical
split-plot dependence is outside the scope of SVEM and is not considered.

The quadratic model has $\pfull = 25$ parameters (including the intercept), so $n_{\text{total}} = \pfull = 25$ is the interpolation boundary. To induce sparsity, in each simulation replicate and for each coefficient $j$, draw $\pi_j \sim \mathrm{Beta}(0.5,0.5)$, $Z_j \sim \mathrm{Bernoulli}(\pi_j)$, and $E_j$ from a mean-zero Laplace distribution, and set $\beta_j = Z_j E_j$. This produces many exact zeros with Laplace-distributed nonzero effects. 

For the Gaussian case, we evaluate the noiseless response surface $f(x)$ on a large
space-filling grid to estimate its domain-scale standard deviation
$\sigma_f=\mathrm{sd}\{f(x)\}$. We then add Gaussian noise to the training
responses using
\[
Y = f(x) + \varepsilon,\qquad \varepsilon\sim\mathcal{N}(0,\sigma_\varepsilon^2),
\qquad
\sigma_\varepsilon = \sigma_f \sqrt{\frac{1-R^2}{R^2}},
\]
so that $\mathrm{Var}\{f(x)\}/\bigl(\mathrm{Var}\{f(x)\}+\mathrm{Var}(\varepsilon)\bigr)\approx R^2$
matches a target $R^2\in\{0.5,0.9\}$.
All methods are scored on a separate noiseless 10{,}000-point holdout.
Performance is summarized by the normalized RMSE,
$\mathrm{NRMSE}=\mathrm{RMSE}/\mathrm{sd}\{f(x)\}$ (with $\mathrm{sd}\{f(x)\}$ computed on the holdout grid),
and we report $\log_e(\mathrm{NRMSE})$.

To probe under- and over-specification, we fit three right-hand-side
expansions against the same data-generating mechanism (whose true order is
quadratic):
\emph{Order 1} = main effects only (underspecified), $\pfull = 7$;
\emph{Order 2} = quadratics and two-way interactions (correct order), $\pfull = 25$; and
\emph{Order 3} = Order~2 plus all three-way interactions and pure cubic terms
(overspecified), $\pfull = 45$. Within each expansion, we compare \SVEMnet objectives
(\wSSE, \wAIC, \wBIC) with and without relaxed base learners, and repeated
$5\times$CV elastic-net baselines using the same $\alpha$ grids.

The binomial study uses the same latent quadratic surface $\eta(x)$ but treats
the binary outcome as a discretized analog. We define
$p(x)=\operatorname{logit}^{-1}\!\big(s\,\eta(x)\big)$ with
$s=\sqrt{R^2/(1-R^2)}$ for $R^2\in\{0.5,0.9\}$, draw training labels
$Y\sim\mathrm{Bernoulli}\{p(x)\}$ once per replicate, and score models on a
large noiseless holdout using the true probabilities $p(x)$. Here $R^2$ serves
as a convenient signal-strength parameter (an approximate analog of the
Gaussian-domain $R^2$ rather than an exact GLM coefficient of determination).
Performance is measured by holdout log-loss. Methods again include
\SVEMnet with \{\wAIC,\ \wBIC,\ \wSSE\} and repeated cross-validated baselines
matched on $\alpha$ and relaxation.

To reduce Monte Carlo error, the summary curves in
Figures~\ref{fig:modelorder}, \ref{fig:kmedian_by_n}, and
\ref{fig:sim_bin_overview} average over 1000 independent simulation
replicates for each combination of $n_{\text{total}}$, target $R^2$,
model order, and fitting method (\texttt{Setting}). In these overview
plots we fix a common set of tuning defaults. For Gaussian responses,
\texttt{cv.glmnet} uses standard (non-relaxed) elastic-net paths
(\texttt{relax = FALSE}) and \SVEMnet{} uses relaxed base learners
(\texttt{relax = TRUE}); post-hoc debiasing is disabled for all methods
and the elastic-net mixing grid is held at $\alpha \in \{0.5, 1\}$.
For binomial responses, all methods use non-relaxed paths
(\texttt{relax = FALSE}) with the same default $\alpha$ grid.
These settings are used for all model orders shown in the figures.

For the mixed-model multiple-comparison summaries in
Tables~\ref{tab:tukey-hsd} and~\ref{tab:tukey-binomial} we focus on the
correctly specified quadratic expansion (Order~2) and explore a richer
grid of SVEM and \texttt{cv.glmnet} configurations. Separate simulation
runs with 250 (Gaussian) and 200 (binomial) independent replicates for
each combination of $n_{\text{total}}$, \texttt{Target\_R2}, and
\texttt{Setting} are used to fit linear mixed models with \texttt{Setting}
as the primary fixed effect, $n_{\text{total}}$, \texttt{Target\_R2}, and
their interaction as additional categorical fixed effects, and a random
intercept for replicate (\texttt{RunID}). Pairwise differences in
log-NRMSE and log-loss are summarized via Tukey--Kramer all-pairs
comparisons of the model-based least-squares means. The loss-only
\wSSE{} selector is included in the simulation runs and overview figures
but excluded from the mixed-model datasets to avoid inflating the number
of clearly noncompetitive comparisons. The resulting connecting-letter
displays appear in Tables~\ref{tab:tukey-hsd} and~\ref{tab:tukey-binomial}.

\subsection{Gaussian results}

Several consistent patterns emerge across $R^2$ levels, sample sizes, and model
orders.

\begin{figure}[!t]
  \centering
  \includegraphics[width=.80\linewidth]{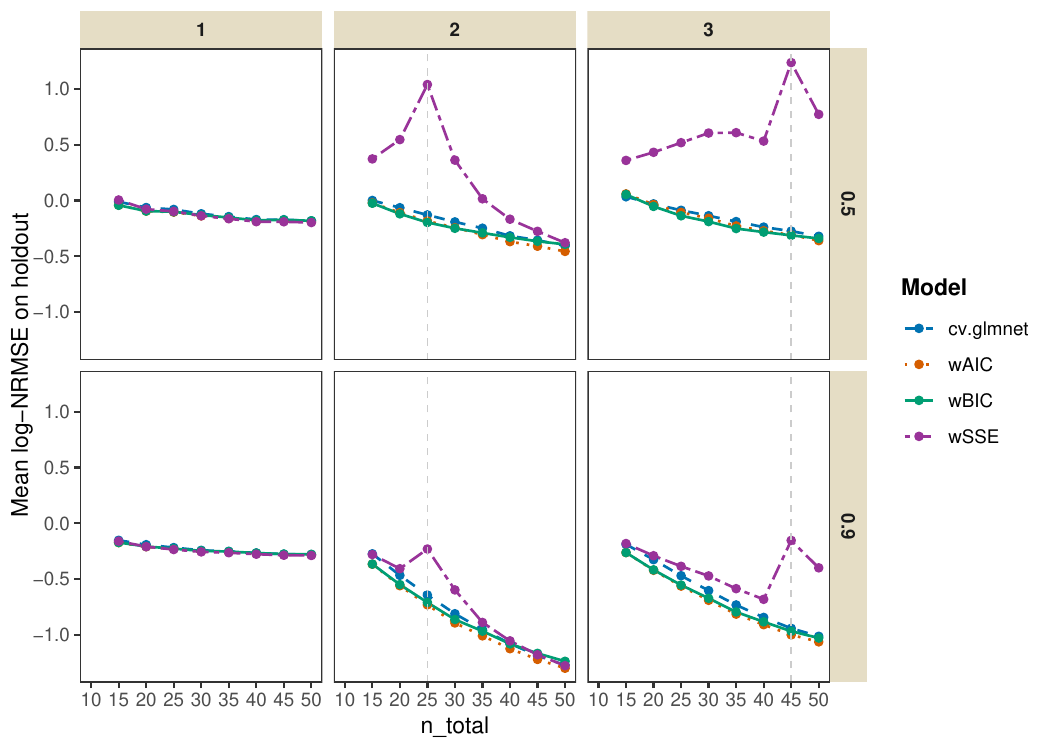}
\caption{Gaussian simulations: mean holdout log-NRMSE (lower is better) versus
$n_{\text{total}}$ by target $R^2$ (rows) and fitted expansion order (columns).
Vertical lines mark interpolation boundaries $n_{\text{total}}=\pfull$.}

  \label{fig:modelorder}
\end{figure}

Figure~\ref{fig:modelorder} summarizes mean log-NRMSE on a large noiseless holdout as a
function of $n_{\text{total}}$ for target $R^2\in\{0.5,0.9\}$ (rows) and three
right-hand-side expansion orders (columns). Curves compare repeated
$5\times$CV elastic-net baselines (\texttt{cv.glmnet}) with \SVEMnet{} selectors
(\wAIC{}, \wBIC{}, and the loss-only \wSSE{}). For these overview plots,
\texttt{cv.glmnet} uses standard (non-relaxed) elastic-net paths,
\SVEMnet{} uses relaxed base learners for Gaussian responses, and post-hoc debiasing is
disabled for all methods.

\begin{figure}[t]
  \centering
  \includegraphics[page=1,width=.80\linewidth]{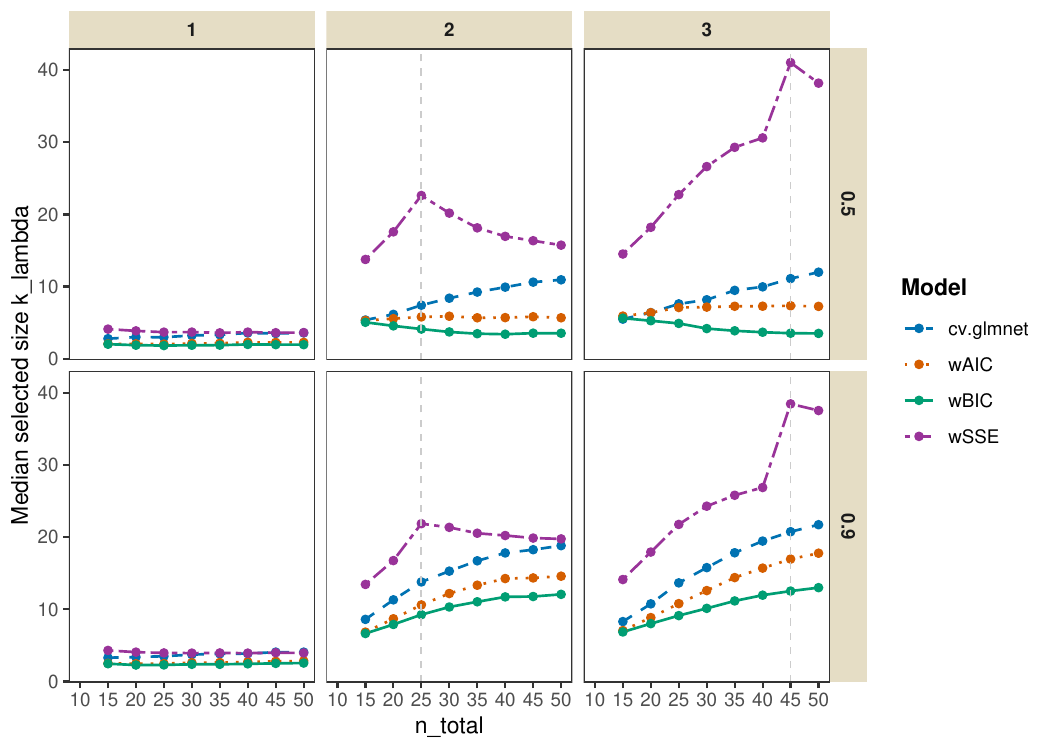}
\caption{Gaussian simulations: mean selected model size versus $n_{\text{total}}$
by target $R^2$ (rows) and expansion order (columns). For SVEM, each training
replicate contributes its median $k_\lambda$ across bootstrap members; CV
contributes its selected model size. Vertical lines mark $n_{\text{total}}=\pfull$.}

  \label{fig:kmedian_by_n}
\end{figure}

To connect predictive error to selected model complexity, Figure~\ref{fig:kmedian_by_n}
reports the mean of the within-replicate median active-set size $k_\lambda$ (number of
nonzero coefficients) selected across SVEM bootstrap replicates. The loss-only
\wSSE{} selector shows a pronounced spike near $n_{\text{total}}=\pfull$,
whereas \wAIC{} and \wBIC{} remain stable.

\begin{itemize}
\item \textbf{Information criteria versus loss-only selectors.}
Across the grid, \SVEMnet{} with validation-weighted information criteria
(\wAIC{}, \wBIC{}) consistently outperforms the loss-only \wSSE{}
selector in mean holdout log-NRMSE. Relative to repeated \texttt{cv.glmnet},
\SVEMnet-\wAIC{} is typically best or tied for best for the correctly specified
and mildly overspecified expansions (Orders~2 and~3), with \wBIC{}
close behind (Figure~\ref{fig:modelorder} and Appendix Table~\ref{tab:tukey-hsd}).

\item \textbf{Relaxation and debiasing.}
For \SVEMnet{} under \wAIC{}/\wBIC{}, enabling relaxed base learners improves
holdout performance in the Gaussian setting, whereas the historical relaxed-CV wrapper has higher error than its
non-relaxed counterpart. The optional post-hoc debiasing step
($y \sim \hat y$ fit on the training data) does not consistently improve
mean holdout error across these settings, so we retain
\texttt{debias = FALSE} as the default for Gaussian responses (Table~\ref{tab:tukey-hsd}).

\item \textbf{Effect of expansion order.}
Underspecification (Order~1, main effects only) substantially degrades
accuracy. The correctly specified quadratic expansion (Order~2) performs best,
and mild overspecification with additional three-way interactions and pure cubic terms
(Order~3) degrades far less than underspecification
(Figure~\ref{fig:modelorder}).

\item \textbf{Peaking near the interpolation boundary.}
For each simulation run we recorded $k_{\mathrm{median}}$, the median number of active parameters $k_\lambda$ across SVEM bootstrap replicates; Figure~\ref{fig:kmedian_by_n} shows the mean of these medians over simulation replicates. The \wSSE selector shows a pronounced spike in both mean log-NRMSE (Figure~\ref{fig:modelorder}) and median
selected model size $k_\lambda$ (Figure~\ref{fig:kmedian_by_n}) when $n_{\text{total}}$ equals the expansion
dimensions: $\pfull = 25$ for Order 2 and $\pfull = 45$ for Order 3 ($\pfull = 7$ for Order 1 is not included in the range of the graph). This is the ``peaking'' phenomenon
near the interpolation boundary in small-sample regression \citep{Kraemer2009LassoPeaking}. In contrast, \wAIC
and \wBIC -- with their effective sample size guardrails -- remain smooth in both error and $k_\lambda$.
\end{itemize}

Overall, the simulations support the default configuration used in
\SVEMnet for Gaussian responses: validation-weighted \wAIC, relaxed elastic-net base learners,
\texttt{debias = FALSE}, and $\alpha\in\{0.5,1\}$.

\subsection{Binomial results}

\begin{figure}[!t]
  \centering
  \includegraphics[page=1,width=.80\linewidth]{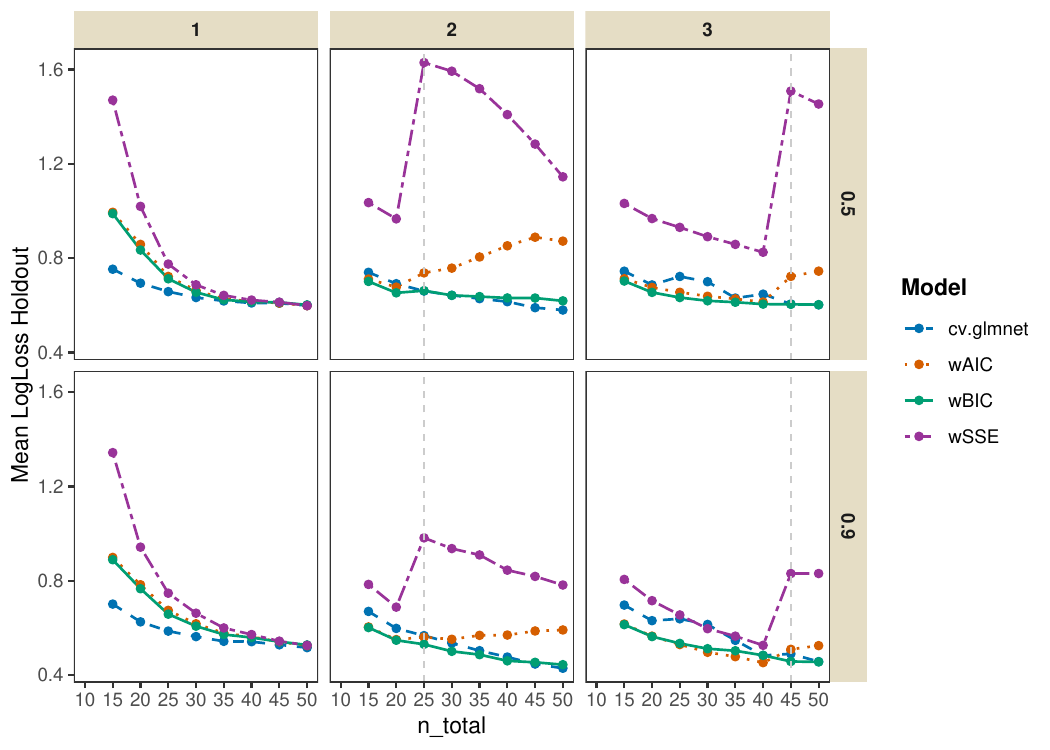}
\caption{Binomial simulations: mean holdout log-loss versus $n_{\text{total}}$ by
target $R^2$ (rows) and fitted expansion order (columns). Vertical lines mark
interpolation boundaries $n_{\text{total}}=\pfull$.}

  \label{fig:sim_bin_overview}
\end{figure}

Across $n_{\text{total}}\in\{15,\dots,50\}$ and $R^2\in\{0.5,0.9\}$,
non-relaxed fits have lower holdout log-loss than their relaxed
counterparts in the reported configurations. For the correctly specified quadratic expansion
(Order~2), \SVEMnet with \wBIC and the non-relaxed repeated-\texttt{cv.glmnet}
baselines perform similarly: Tukey--Kramer comparisons based on the mixed-effects model
place these configurations in the same best-performing group
(Table~\ref{tab:tukey-binomial}). However, Figure~\ref{fig:sim_bin_overview} suggests that \SVEMnet with \wBIC may outperform \texttt{cv.glmnet} in the overspecified case.

Within these historical implementations, relaxed refits have higher error: for each objective and
$\alpha$ grid the \texttt{relax = TRUE} versions have larger
least-squares means, with the relaxed cross-validated baselines forming
the worst group in Table~\ref{tab:tukey-binomial}. The degradation from
relaxation is especially pronounced for the repeated-\texttt{cv.glmnet}
fits, where the increase in mean log-loss is about two to three times as large as
for the corresponding \SVEMnet selectors.

Allowing $\alpha\in\{0.5,1\}$ has no material impact relative to
lasso-only fits ($\alpha=1$): default two-point grids and pure lasso
grids are indistinguishable within each selection method. As in the
Gaussian case, the loss-only \wSSE{} selector (included only in the
overview plots) exhibits a pronounced error spike at
$n_{\text{total}}=\pfull$, now on the log-loss scale
(Figure~\ref{fig:sim_bin_overview}). \SVEMnet with \wAIC underperforms
\wBIC in the binomial setting. These
simulations support the default configuration used in \SVEMnet for
binomial responses: validation-weighted \wBIC, no relaxation, and
$\alpha\in\{0.5,1\}$.

Figure~\ref{fig:sim_bin_overview} summarizes the corresponding mean holdout log-loss
curves for the overview settings (all methods non-relaxed; default $\alpha$ grid).

\section{Conclusion}

\SVEMnet{} provides an open-source implementation of Self-Validated Ensemble
Models for small-sample DOE and chemometric applications, with (relaxed)
elastic-net base learners for Gaussian and binomial responses built on
\glmnet{} using validation-weighted information-criterion analogs. Together with the permutation whole-model test and the random-search
desirability scoring and candidate-selection tools, \SVEMnet{} offers an
end-to-end workflow for mixture-constrained multi-response optimization in a single
toolchain. The software is aimed at practitioners running small-sample
(\emph{e.g.}, tens of runs) chemometric, pharmaceutical, and industrial DOE studies for
formulation optimization, often involving mixture factors alongside numeric and
categorical process factors.

\SVEMnet{} is designed for single-error-stratum settings in which runs can be treated
as independent resampling units. Hierarchical split-plot and other multi-stratum
designs with random effects are not supported by SVEM and would require substantial methodological changes beyond the scope
of this paper. In addition, the validation-weighted information-criterion analogs,
WMT ``$p$-values,'' and mean-in-specification probabilities are intended as heuristic
tuning, ranking, and diagnostic summaries rather than as fully calibrated inferential
quantities. We therefore recommend reporting key tuning budgets and performing simple
sensitivity checks (e.g., varying seeds, bootstrap/permutation budgets, desirability
parameters, or expansion order) when conclusions are marginal. Important
recipes selected by optimization warrant independent confirmation, because
the pointwise uncertainty summaries provide neither simultaneous nor
selection-adjusted coverage guarantees.
\appendix

\section{Validation-weighted criterion and guardrails}\label{app:criterion}

This appendix records the validation-weighted selection criteria used in
\SVEMnet.

\subsection{Gaussian criterion}

Within each FRW replicate, let $\lambda$ index a point on the elastic-net
path, let $k_\lambda$ count the intercept plus the nonzero slope
coefficients, and let $w_i^{\mathrm{valid}}$ be the FRW validation weights,
rescaled to have mean one so that $\sum_i w_i^{\mathrm{valid}} = n$. Write
$r_{\lambda,i} = y_i - \hat y_{\lambda,i}$ for the residual at observation $i$
under~$\lambda$, and define the weighted sum of squared errors
\begin{equation}
\label{eq:ssew}
\mathrm{SSE}_w(\lambda)
=
\sum_i w_i^{\mathrm{valid}} r_{\lambda,i}^2.
\end{equation}

For Gaussian responses we consider three selectors: a loss-only rule based on
$\mathrm{SSE}_w(\lambda)$ and two information-criterion analogs. The
loss-only selector \wSSE{} minimizes $\mathrm{SSE}_w(\lambda)$ over the path.
The validation-weighted AIC and BIC analogs minimize
\begin{equation}
\label{eq:crit}
C_g(\lambda)
=
n \log\!\left(
    \frac{\mathrm{SSE}_w(\lambda)}{\,n\,}
  \right)
+ g\,k_\lambda,
\qquad
g\in\{2,\ \log(n_{\mathrm{eff}}^{\mathrm{adm}})\},
\end{equation}
which yield, respectively, a validation-weighted AIC analog (\wAIC, $g=2$)
and a validation-weighted BIC analog (\wBIC, $g=\log(n_{\mathrm{eff}}^{\mathrm{adm}})$)
\citep{Akaike1974,Schwarz1978,BurnhamAnderson2002}. In the unweighted case,
$w_i^{\mathrm{valid}}\equiv 1$ and $\mathrm{SSE}_w(\lambda)$ reduces to the
usual residual sum of squares $\mathrm{RSS}(\lambda)$. In that case
$C_g(\lambda)$ reduces to $n\log\{\mathrm{RSS}(\lambda)/n\}+g\,k_\lambda$,
which is the standard Gaussian AIC/BIC form (up to an additive constant
independent of~$\lambda$).

Following \citet{Kish1965}, we summarize the variability of the validation
weights via an effective sample size
\begin{equation}
\label{eq:neff}
n_{\mathrm{eff}}
=
\frac{\bigl(\sum_i w_i^{\mathrm{valid}}\bigr)^2}{\sum_i (w_i^{\mathrm{valid}})^2}
\end{equation}
and use the bounded quantity
\begin{equation}
\label{eq:neffadm}
n_{\mathrm{eff}}^{\mathrm{adm}}
=
\min\{\,n,\;\max(2,\;n_{\mathrm{eff}})\,\}
\end{equation}
when forming the BIC-style penalty $\log(n_{\mathrm{eff}}^{\mathrm{adm}})\,k_\lambda$.
With mean-one FRW weights, $2 \le n_{\mathrm{eff}}^{\mathrm{adm}} \le n$.

For the information-criterion selectors (\wAIC, \wBIC) we additionally impose
\begin{equation}
\label{eq:guardrail}
k_\lambda - 1 \;<\; n_{\mathrm{eff}}^{\mathrm{adm}},
\end{equation}
so that the number of non-intercept coefficients is strictly smaller than the
effective number of validation observations. Path points violating this
inequality are treated as inadmissible when evaluating $C_g(\lambda)$.
This guardrail helps stabilize model selection near the interpolation boundary
$n=\pfull$ \citep{Efron2004LARS,ESL2009,Kraemer2009LassoPeaking,Belkin2019DoubleDescent}.

\subsection{Binomial (logistic) criterion}

For binomial responses we minimize a deviance-style quantity plus a complexity
penalty. Let $\ell_{\mathrm{valid}}(\lambda)$ denote the weighted
log-likelihood on the FRW validation copy under path point $\lambda$, and let
$\mathrm{NLL}(\lambda) = -\ell_{\mathrm{valid}}(\lambda)$ be the corresponding
weighted negative log-likelihood. The AIC- and BIC-style selectors minimize
\begin{equation}
\label{eq:crit_bin}
C_g^{(\mathrm{bin})}(\lambda)
=
-2\,\ell_{\mathrm{valid}}(\lambda)\;+\;g\,k_\lambda
=
2\,\mathrm{NLL}(\lambda)\;+\;g\,k_\lambda,
\qquad
g\in\{2,\,\log(n_{\mathrm{eff}}^{\mathrm{adm}})\}.
\end{equation}
The loss term $-2\ell_{\mathrm{valid}}(\lambda)$ differs from the weighted
binomial deviance only by an additive constant independent of $\lambda$. In the implementation we work directly with
$\mathrm{NLL}(\lambda)$ and form $C_g^{(\mathrm{bin})}(\lambda)$ as
$2\times\text{NLL} + g\,k_\lambda$ for these selectors. For \wAIC{} and
\wBIC{} the same effective-size quantity $n_{\mathrm{eff}}^{\mathrm{adm}}$ and
guardrail $k_\lambda-1 < n_{\mathrm{eff}}^{\mathrm{adm}}$ are used as in the
Gaussian case.

When $g=0$ the selector is a loss-only rule based on the weighted negative
log-likelihood; for consistency with the Gaussian setting we retain the
label \wSSE{} for this case even though it is based on negative
log-likelihood rather than squared error. Ensemble predictions are formed by
averaging bootstrap-member-predicted probabilities on the response scale.

\section{Supplementary robustness and diagnostic analyses}\label{app:supp}

This appendix provides  additional robustness checks and diagnostics. All simulation scripts are available in the
Mendeley Data repository 
\citep{Karl2025_SVEMnetData}.

\subsection{Focused interpolation-boundary sweep}\label{app:supp_boundary}

To run an explicit check of boundary behavior near the interpolation point,
we ran a fine sweep of the total run size $n$ in a representative Gaussian setting,
drawing a new true response surface and a new (noiseless) evaluation set in each replicate.
The deterministic expansion order was fixed at Order~2, with $p_{\mathrm{full}}=25$ coefficients (including the intercept,
main effects, two-way interactions, and quadratic terms). We varied $n$ over
$\{20,21,\ldots,30\}$, spanning the interpolation boundary $n \approx p_{\mathrm{full}}$,
and used $150$ replicates per $n$ (new training designs and noise per replicate) with a target strength of $R^2=0.5$.

\begin{figure}[t]
  \centering
  \includegraphics[width=\linewidth]{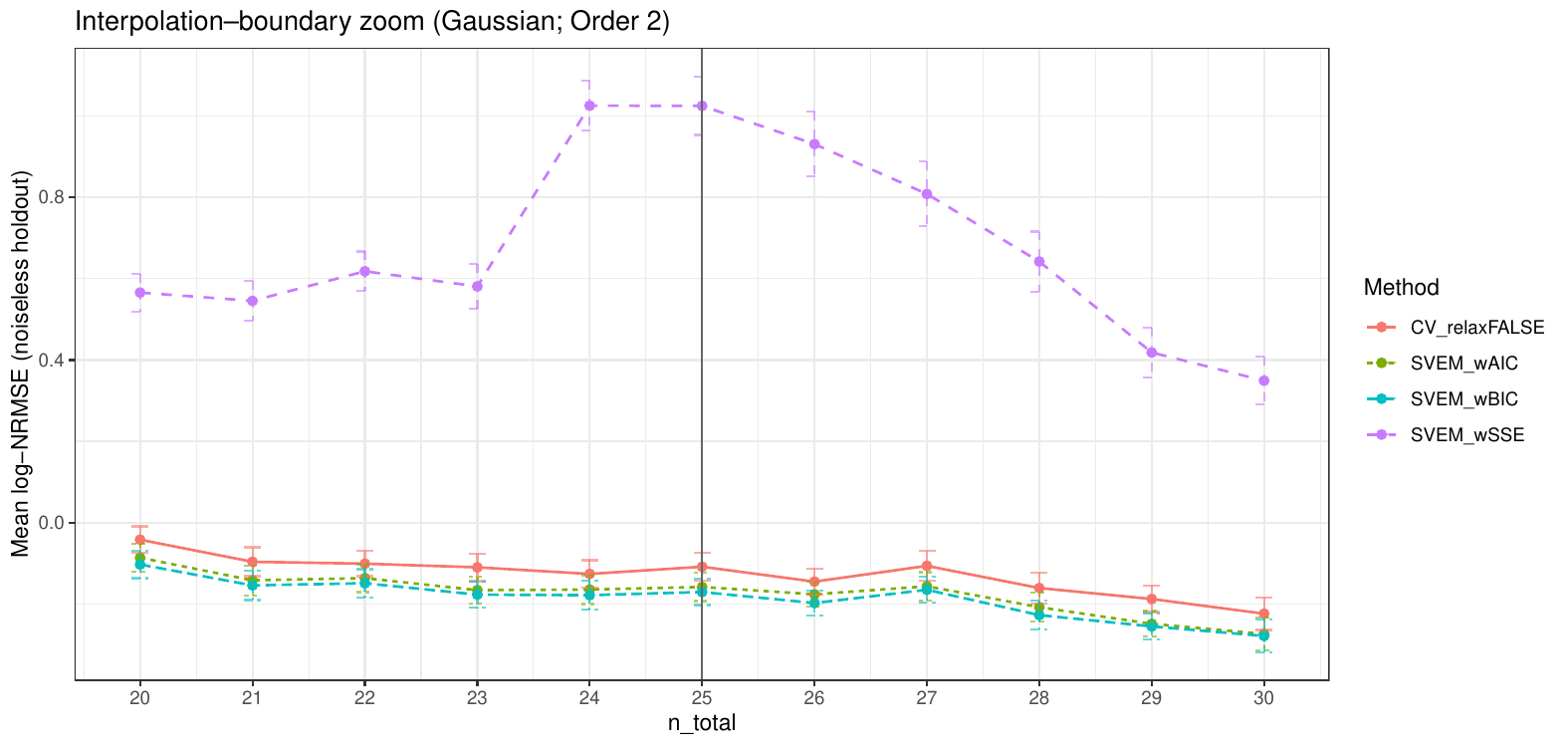}\\[-2pt]
  \includegraphics[width=\linewidth]{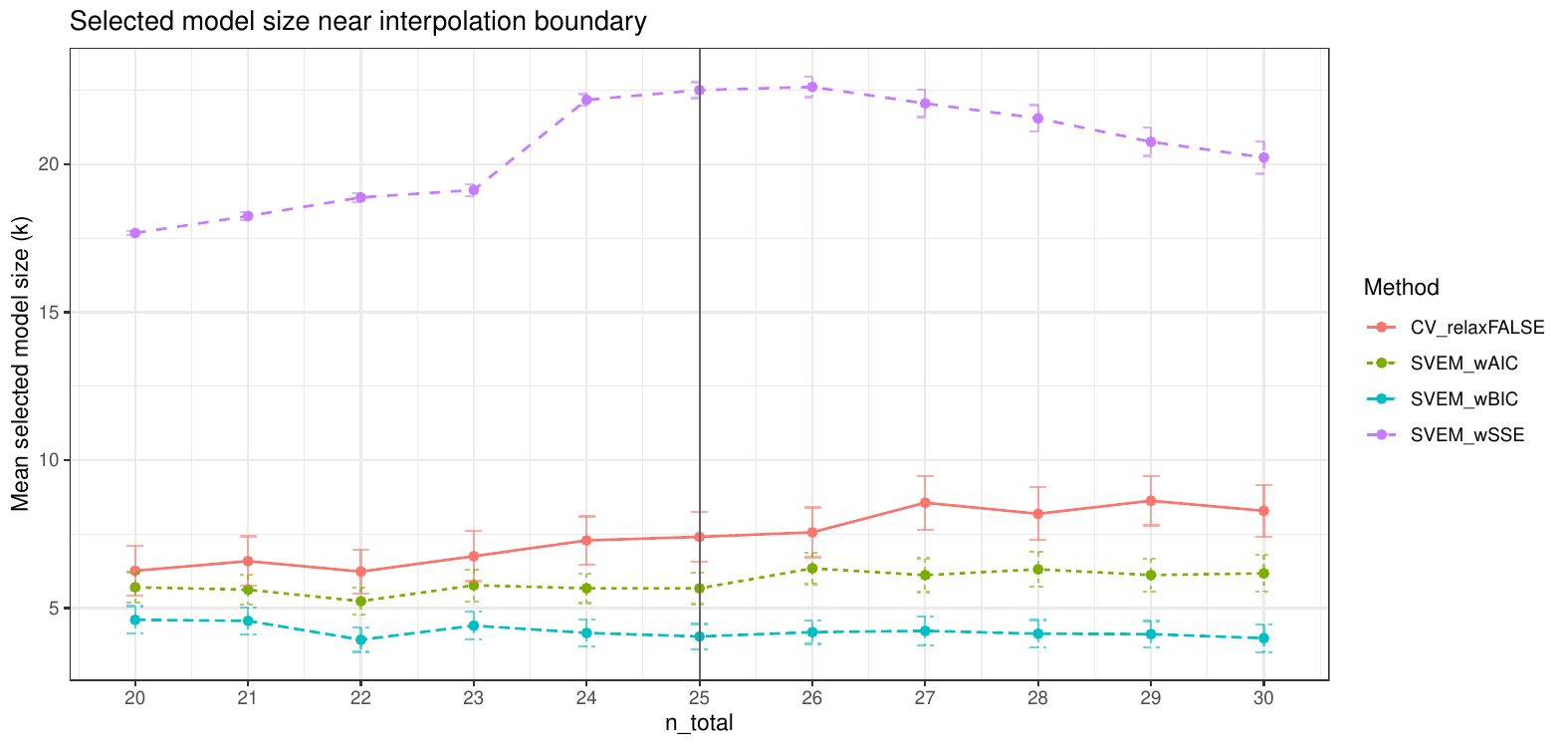}
  \caption{Focused interpolation-boundary sweep (Gaussian). Top: mean holdout log-NRMSE
  versus $n$ near $p_{\mathrm{full}}=25$. Bottom: mean selected model size versus $n$ (within-replicate median
  $k_\lambda$ for SVEM; selected model size for CV).}
  \label{fig:supp_boundary}
\end{figure}

Figure~\ref{fig:supp_boundary} shows the mean holdout log-NRMSE and the mean of the
per-replicate median selected model size $k_\lambda$ as a function of $n$. The loss-only selector \wSSE{} exhibits
a pronounced ``peaking''/instability near the boundary (worst at $n=24$ and $n=25$),
while \wAIC{} and \wBIC{} remain stable and avoid selecting near-saturated
models. At the boundary $n=25$, the mean holdout NRMSE under
\wSSE{} was 3.07, compared to 0.87 and 0.86 for
\wAIC{} and \wBIC{}, respectively (repeated \texttt{cv.glmnet} baseline:
0.92). Consistent with the manuscript discussion, \wAIC/\wBIC{} act as
validation-weighted tuning-score analogs under FRW rather than classical information
criteria with formal asymptotic justification; empirically, they suppress boundary peaking by penalizing
overly large effective model size within each FRW replicate.

\subsection{Design robustness and effect-density/distribution robustness}\label{app:supp_ccd}

To qualify the simulation evidence beyond the Latin-hypercube baseline and sparse-effect
settings, we repeated a representative Gaussian simulation under two design regimes:
(i) a Latin hypercube design (LHS; baseline in the paper) and
(ii) a face-centered central composite design (CCD), matched to the same factor space.
For computational simplicity in this robustness check, we restrict attention to the
four continuous factors $X_1{:}X_4\in[-1,1]$ (no categorical factor), and we fit the same
quadratic (Order~2) deterministic expansion used in the main Gaussian study.

The CCD uses the standard face-centered structure for $k=4$ continuous factors:
$2^4$ factorial points plus $2k$ axial (face) points, augmented with two center points,
for a total of $n_{\text{total}}=2^4+2k+2=24+2=26$ runs. The LHS design is generated on the same
$[-1,1]^4$ domain with $n_{\text{total}}=26$ runs.

To study sparsity sensitivity, we used three coefficient settings for the true surface:
a sparse Laplace regime (baseline), a sparse Gaussian regime, and a denser Gaussian regime,
each evaluated at two target signal strengths ($R^2\in\{0.5,0.9\}$). Each
design--effect--$R^2$ setting used $500$ independent replicates (new coefficient vector, training data, and noise per replicate; the LHS design is regenerated each replicate, while the CCD uses the fixed face-centered point set with randomized run order),
and all methods were scored on a large noiseless holdout set.

\begin{figure}[t]
\centering
\includegraphics[width=\linewidth]{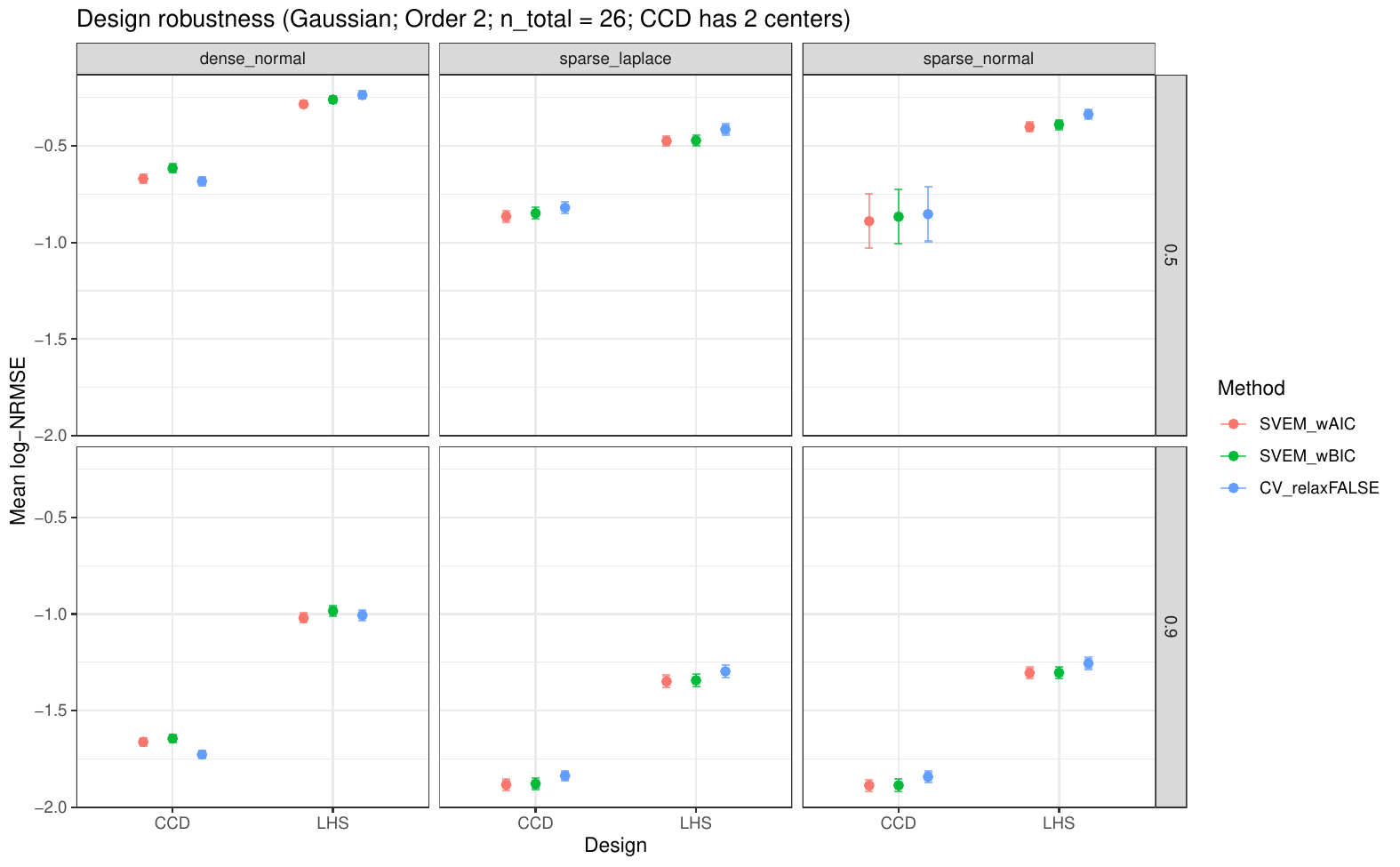}
\caption{Design robustness and effect-density/distribution robustness (Gaussian; continuous factors $X_1{:}X_4$ only; $n_{\text{total}}=26$; Order~2).
Mean holdout log-NRMSE (95\% Monte Carlo standard-error intervals) comparing LHS and face-centered CCD designs across three true-surface regimes (sparse Laplace, sparse Normal, dense Normal)
and two target signal levels ($R^2\in\{0.5,0.9\}$). The CCD uses the standard face-centered $k=4$ structure ($2^4+2k$) with two center points.}
\label{fig:supp_ccd_dense}
\end{figure}

Figure~\ref{fig:supp_ccd_dense} summarizes mean holdout log-NRMSE (with 95\% Monte Carlo standard-error intervals).
In this study the CCD yields uniformly lower error than the matched LHS design across all settings and
methods, which is expected because the data-generating surface is quadratic and therefore closely aligned
with the model class that CCDs are designed to estimate efficiently. Within this (best-case) CCD regime,
\wAIC and \wBIC remain competitive with repeated \texttt{cv.glmnet}; however,
for the CCD with a dense-Normal coefficient regime, repeated CV is better, most clearly at
high signal ($R^2=0.9$), and only very slightly at $R^2=0.5$.

More broadly, these results should be interpreted as complementary to (not a replacement for) the paper's baseline space-filling evidence: in formulation optimization one often fits a large expansion with many candidate interactions and polynomial terms to hedge against surface complexity, while expecting that only a small subset of terms is active (the sparsity-of-effects principle). Highly structured classical designs can
perform extremely well when the assumed model class is correct, but when the active terms are not well supported by the design (or when the working expansion is misspecified relative to the true surface),
the resulting model can be biased and predictive performance can degrade \citep{KarlEtAl2022_JMPMixture}. A systematic study of
design choice under misspecification and varying sparsity---and how \SVEMnet\ and repeated CV behave
in those regimes---is a natural direction for future work, and the released \SVEMnet\ package and
shared simulation code make such extensions straightforward to implement.

\subsection{Relaxation diagnostics}\label{app:supp_relax}

\paragraph{Gaussian}
To make the bias--variance and relaxation behavior explicit, we ran a representative Gaussian diagnostic with a fixed true surface (sparse-Laplace coefficients) and a fixed noiseless holdout set. We re-sampled the training design and noise across $N_{\mathrm{rep}}=500$ replicates using an Order~2 expansion with $n_{\text{total}}=30$ and target $R^2=0.9$. We compared four methods:
(i) \wAIC with relaxation enabled,
(ii) \wAIC with relaxation disabled,
(iii) a repeated \texttt{cv.glmnet} baseline with relaxation disabled, and
(iv) a repeated \texttt{cv.glmnet} baseline with relaxation enabled.
To reduce tuning variability in this diagnostic and to keep $\lambda$ comparisons interpretable, all fits use a lasso-only grid ($\alpha=1$).

For each method, we recorded holdout NRMSE/log-NRMSE and a prediction-scale decomposition of holdout mean-squared error into $\mathrm{Bias}^2$ and $\mathrm{Var}$ components, defined as
$\mathrm{Bias}^2 = \mathbb{E}_{x}\{(\mathbb{E}_{r}\hat{y}_{r}(x) - y(x))^2\}$ and
$\mathrm{Var} = \mathbb{E}_{x}\{\mathrm{Var}_{r}(\hat{y}_{r}(x))\}$,
where $r$ indexes training replicates and expectations average over holdout points.
We additionally recorded, per replicate, the selected regularization level $\lambda$
(for \SVEMnet, summarized as the median of $\log_{10}(\lambda)$ across bootstrap-selected path points; for repeated CV, the single selected $\log_{10}(\lambda)$).

\begin{figure}[!htbp]
  \centering
  \includegraphics[width=\linewidth,height=.34\textheight,keepaspectratio]{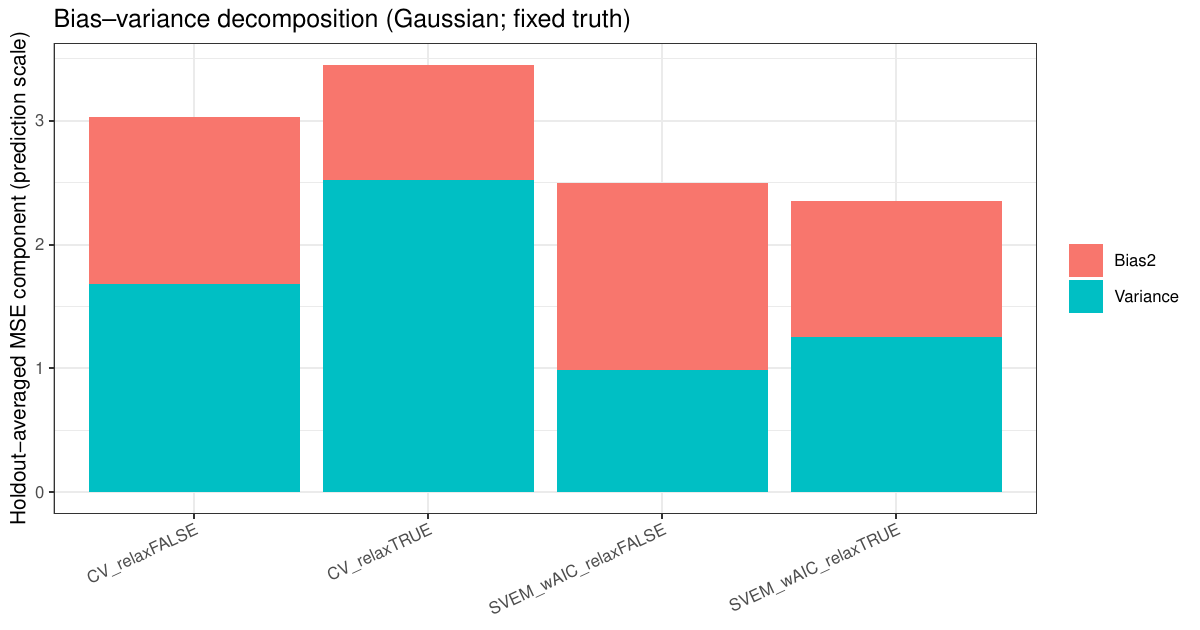}\\[-2pt]
  \includegraphics[width=\linewidth,height=.34\textheight,keepaspectratio]{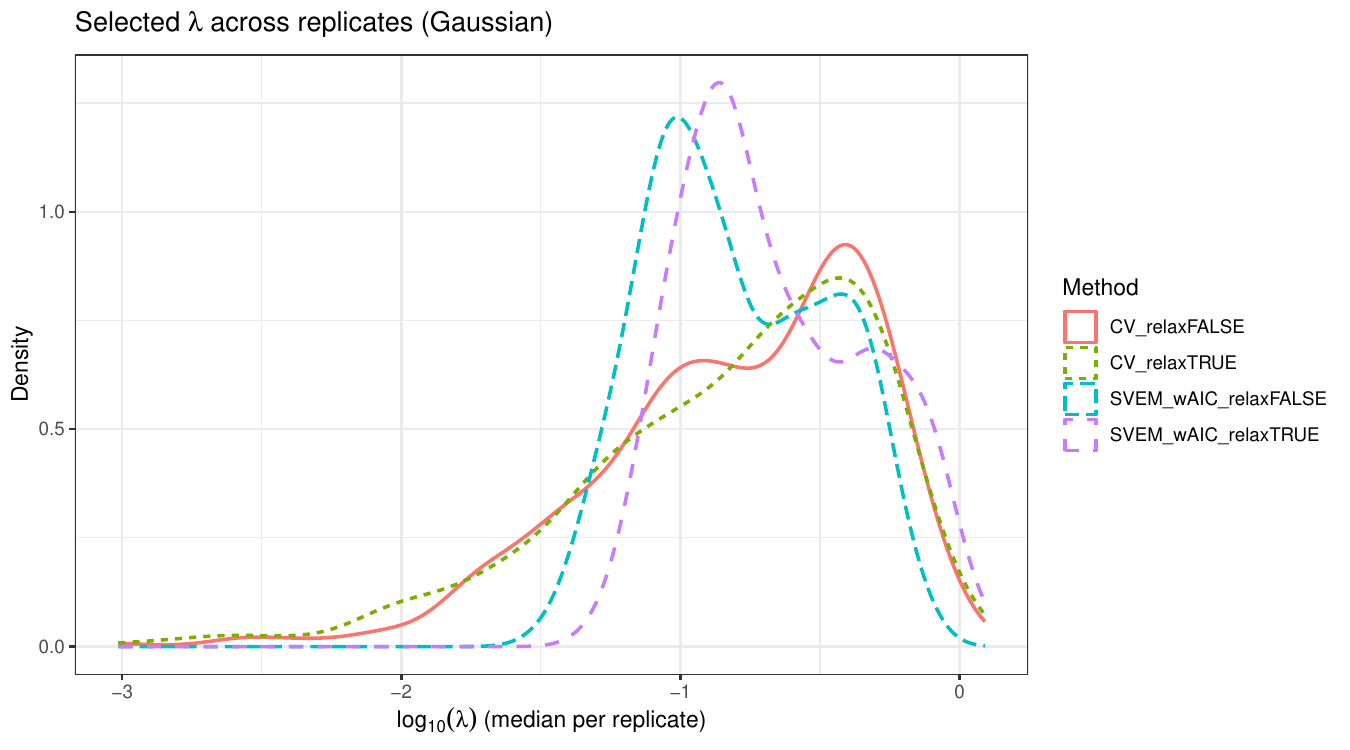}
  \caption{Gaussian relaxation diagnostics ($n_{\text{total}}=30$, Order~2, $N_{\mathrm{rep}}=500$, lasso-only $\alpha=1$).
  Top: prediction-scale bias--variance decomposition on a fixed noiseless holdout set.
  Bottom: distribution of selected $\log_{10}(\lambda)$ across training replicates (per replicate: median across SVEM bootstraps; single selected value for repeated CV).}
  \label{fig:supp_relax_gaussian}
\end{figure}

Figure~\ref{fig:supp_relax_gaussian} shows the resulting bias--variance decomposition and the distribution of selected $\log_{10}(\lambda)$ across replicates. In this setting, relaxation exhibits the expected trade-off: it lowers bias but increases variance. For \wAIC, relaxation reduces $\mathrm{Bias}^2$ (1.10 vs 1.51) while increasing $\mathrm{Var}$ (1.25 vs 0.99), yielding a small improvement in overall MSE (2.35 vs 2.50) and mean holdout NRMSE (0.439 vs 0.454). For the historical repeated-CV wrapper, the relaxed setting lowers bias (0.93 vs 1.35) but increases variance more sharply (2.52 vs 1.68), worsening overall MSE (3.45 vs 3.03) and mean NRMSE (0.528 vs 0.499). In the relaxed \SVEMnet\ fits, the strongest relaxation available on the fixed grid was selected throughout ($\gamma=0.2$), so we omit a separate $\gamma$ plot and focus on the induced changes in prediction error and the $\lambda$ distribution. To keep the relaxed search computationally tractable, \SVEMnet\ evaluates a coarse fixed grid $\gamma\in\{0.2,0.6,1.0\}$ within each bootstrap replicate, where $\gamma=1$ corresponds to the standard (non-relaxed) elastic-net fit at the selected $\lambda$, and smaller $\gamma$ values apply a less-shrunk refit on the same active set (with the limiting case $\gamma\to 0$ approaching an unpenalized refit on the selected active predictors).

\paragraph{Binomial}
We repeated an analogous diagnostic under a representative binomial setting ($n=25$, Order~2,
target $R^2=0.5$, $N_{\mathrm{rep}}=200$), comparing \wBIC{} and repeated
\texttt{cv.glmnet} with relaxation enabled/disabled. The binomial case reinforces the
recommendation in the main text to disable relaxation: in this run, relaxation increased
prediction variance and worsened holdout log-loss for both historical
implementations; the repeated-CV caveat in Section~\ref{sec:sim} also applies here.
Mean log-loss was 0.345 for non-relaxed \wBIC{}
and 0.347 for non-relaxed repeated CV, versus
0.371 and 0.362 for the
relaxed versions.

\subsection{LNP example: WMT stability over random seeds and permutation budgets}\label{app:supp_wmt}

Because WMT is a Monte Carlo diagnostic, we performed a small stability check for the
LNP example by repeating \lstinline|svem_wmt_multi()| under five random seeds and two
permutation budgets ($n_{\mathrm{perm}}=150$ (default) and $n_{\mathrm{perm}}=500$),
holding all other settings fixed (shared deterministic expansion and mixture constraints).
Table~\ref{tab:supp_wmt_stability} summarizes the approximate whole-model $p$-values and the
corresponding WMT multipliers.

These $p$-values are produced by the same parametric-tail procedure used in the package:
WMT fits a flexible SHASH distribution to the permutation distance distribution and then reports
the fitted right-tail probability for the observed distance. Because this is an extrapolated tail
estimate rather than the raw empirical permutation fraction, values can be much smaller than
$1/(n_{\mathrm{perm}}+1)$; we therefore interpret them as approximate diagnostic probabilities
rather than exact permutation-test $p$-values.

Across seeds and budgets, Size and Potency consistently show strong global signal
(always $p \le \num{4.45e-7}$, often at the numerical floor), whereas PDI remains
weak (here $p$ in the range \num{0.76}--\num{0.94}). Increasing $n_{\mathrm{perm}}$ narrows
the Potency $p$-values but does not change the qualitative conclusion.

\begin{table}[t]
\centering
\scriptsize
\caption{WMT stability check for the LNP example. For each response and permutation budget
($n_{\mathrm{perm}}$), values shown are median [min, max] across five random seeds.}
\label{tab:supp_wmt_stability}
\begin{tabular}{l c c c}
\toprule
Response & $n_{\mathrm{perm}}$ & $p$ (median [min,max]) & multiplier (median [min,max])\\
\midrule
Size & 150 & $1\mathrm{e}{-16}$ [$1\mathrm{e}{-16}$, $1\mathrm{e}{-16}$] & 16 [16, 16]\\
Potency & 150 & $1\mathrm{e}{-16}$ [$1\mathrm{e}{-16}$, $4.45\mathrm{e}{-7}$] & 16 [6.35, 16]\\
PDI & 150 & $0.832$ [$0.765$, $0.935$] & 0.08 [0.029, 0.12]\\
\midrule
Size & 500 & $1\mathrm{e}{-16}$ [$1\mathrm{e}{-16}$, $1\mathrm{e}{-16}$] & 16 [16, 16]\\
Potency & 500 & $2.23\mathrm{e}{-11}$ [$5.42\mathrm{e}{-13}$, $6.86\mathrm{e}{-8}$] & 10.7 [7.16, 12.3]\\
PDI & 500 & $0.833$ [$0.783$, $0.9$] & 0.079 [0.046, 0.11]\\
\bottomrule
\end{tabular}
\end{table}
\subsection{LNP example: desirability calibration sensitivity}\label{app:supp_desirability}

To operationalize desirability calibration and provide a lightweight sensitivity check, we held the
fitted SVEM surrogate models fixed and re-scored random-search LNP candidates under several small,
practitioner-relevant perturbations of the desirability mapping. We repeated this across five
independent random candidate draws ($N=25{,}000$ candidates per draw, with the same mixture constraints)
and, within each draw, compared the resulting top-$K$ candidate sets against the default mapping using
Jaccard overlap. We report results for $K\in\{50,200\}$, where $K=200$ provides a broad stability summary
and $K=50$ reflects a more stringent ``top list''.

In this sensitivity exercise, ``acceptable bounds'' refer to the per-response Derringer--Suich
mapping parameters that define the onset of full desirability (e.g., a lower acceptable value for
a ``maximize'' goal, or an upper acceptable value for a ``minimize'' goal). Under the default mapping,
these acceptable bounds are inferred automatically from the predicted response distributions over the
candidate set within each draw (using a quantile-anchored range with a small span floor for stability;
see code in the Mendeley repository \citep{Karl2025_SVEMnetData} for details). Variant
\texttt{V4\_bounds\_tighten5} then tightens these inferred acceptable bounds by $5\%$ of the inferred span
(within the same draw), representing a mild perturbation around the default anchoring. In contrast,
\texttt{V6\_spec\_bounds\_alt} replaces the inferred acceptable bounds with fixed, practitioner-chosen
``specification-anchored'' ranges (Potency: 78--95; Size: 50--100; PDI: 0.10--0.25), while keeping the
same goals and weights; this isolates sensitivity to how acceptable ranges are anchored rather than to
small local perturbations around a fixed anchoring.

Tables~\ref{tab:supp_desirability_k50}--\ref{tab:supp_desirability_k200} show that mild perturbations to
objective weights, a single-response shape exponent, or modest tightening of the inferred acceptable bounds
yield high stability of the top set. Uniformly scaling all shape exponents leaves the top-$K$ set unchanged.
In contrast, the specification-anchored calibration produces a partially different top set, consistent with
recommendations being more sensitive to how practitioners anchor acceptable ranges than to modest
curvature/weight perturbations around a fixed anchoring.

\begin{table}[t]
\centering
\caption{Desirability sensitivity (LNP example), $K=50$. Top-$K$ overlap vs.\ the default desirability mapping
under small perturbations. Entries are median [min,max] over five independent random candidate draws
($N=25{,}000$ each). Overlap reports $|A\cap B|$ out of $K$.}
\label{tab:supp_desirability_k50}
\footnotesize
\setlength{\tabcolsep}{5pt}
\renewcommand{\arraystretch}{1.15}
\begin{tabular}{llcc}
\hline
Variant & Description (brief) & $|A\cap B|$ & Jaccard \\
\hline
V1\_uniform\_shape2\_sanity & Uniform DS shape $=2$ (sanity check)
  & 50 [50,50] & 1.000 [1.000,1.000] \\
V2\_weights\_pot\_mild & Mild reweight toward potency (0.65/0.25/0.10)
  & 47 [46,48] & 0.887 [0.852,0.923] \\
V3\_size\_shape125 & Size desirability shape $=1.25$ (mildly steeper)
  & 46 [42,48] & 0.852 [0.724,0.923] \\
V4\_bounds\_tighten5 & Tighten inferred acceptables by 5\% of span
  & 50 [50,50] & 1.000 [1.000,1.000] \\
V6\_spec\_bounds\_alt & Spec-anchored acceptable bounds (alternative)
  & 32 [28,37] & 0.471 [0.389,0.587] \\
\hline
\end{tabular}
\end{table}

\begin{table}[t]
\centering
\caption{Desirability sensitivity (LNP example), $K=200$. Top-$K$ overlap vs.\ the default desirability mapping
under small perturbations. Entries are median [min,max] over five independent random candidate draws
($N=25{,}000$ each). Overlap reports $|A\cap B|$ out of $K$.}
\label{tab:supp_desirability_k200}
\footnotesize
\setlength{\tabcolsep}{5pt}
\renewcommand{\arraystretch}{1.15}
\begin{tabular}{llcc}
\hline
Variant & Description (brief) & $|A\cap B|$ & Jaccard \\
\hline
V1\_uniform\_shape2\_sanity & Uniform DS shape $=2$ (sanity check)
  & 200 [200,200] & 1.000 [1.000,1.000] \\
V2\_weights\_pot\_mild & Mild reweight toward potency (0.65/0.25/0.10)
  & 187 [185,188] & 0.878 [0.860,0.887] \\
V3\_size\_shape125 & Size desirability shape $=1.25$ (mildly steeper)
  & 191 [186,193] & 0.914 [0.869,0.932] \\
V4\_bounds\_tighten5 & Tighten inferred acceptables by 5\% of span
  & 200 [199,200] & 1.000 [0.990,1.000] \\
V6\_spec\_bounds\_alt & Spec-anchored acceptable bounds (alternative)
  & 168 [159,168] & 0.724 [0.660,0.724] \\
\hline
\end{tabular}
\end{table}

\subsection{LNP example: candidate stability under adjacent expansion orders}\label{app:supp_modelorder}

To demonstrate stability of candidate sets under adjacent expansion-order choices, we fit two adjacent deterministic expansions (Orders~2 and~3) to the
same LNP training data and compared the top-$K$ candidate sets ($K=200$) under the primary score.
The overlap was $155/200$ candidates (Jaccard $=0.633$), indicating moderate stability:
many top candidates persist across orders, but some candidates shift ranking when the expansion
is changed. In practice, this motivates the lightweight order-sensitivity check recommended in
the main text.

\section{Supplementary multiple-comparison tables}\label{app:tukey}

For completeness we provide Tukey--Kramer multiple-comparison summaries
for the Gaussian and binomial simulations. These tables correspond to
the mixed-model analyses described in Section~\ref{sec:sim}, based on
250 (Gaussian) and 200 (binomial) independent simulation replicates for
each combination of $n_{\text{total}}$, \texttt{Target\_R2}, and
\texttt{Setting} at the quadratic expansion (Order~2); full CSV outputs
are included in the replication bundle \citep{Karl2025_SVEMnetData}.

For all Tukey--Kramer analyses we fixed the expansion at the correctly specified
quadratic order (Order~2) and fit linear mixed models in JMP Pro 19 with \texttt{Setting} as
the primary factor of interest, $n_{\text{total}}$, \texttt{Target\_R2}, and
their interaction as categorical fixed effects, and simulation run ID as a
random intercept. Pairwise differences between settings were assessed using
Tukey--Kramer honestly significant difference comparisons of the model-based
least-squares means, controlling the familywise error rate at
$\alpha = 0.05$. Settings that do not share a letter differ significantly, and smaller
least-squares means correspond to better performance (lower log-NRMSE or
log-loss). Here \texttt{relaxTRUE/FALSE} indicates whether relaxed base
learners are used, \texttt{lasso/default} indicates the pure lasso case
($\alpha = 1$) versus the mixed-$\alpha$ grid $\alpha \in \{0.5, 1\}$, and in
Table~\ref{tab:tukey-hsd} the suffix \texttt{dbTRUE/FALSE} indicates whether
the post-hoc debiasing step is applied. In Table~\ref{tab:tukey-binomial} all
configurations use \texttt{dbFALSE}. The \texttt{CV\_relaxTRUE} rows use
the historical 3.2.0 relaxed-CV wrapper described in Section~\ref{sec:sim}.

To avoid inflating the number of pairwise comparisons with a clearly
noncompetitive method, the loss-only \wSSE{} selector was omitted from the
Tukey--Kramer analyses. The overview plots in
Figures~\ref{fig:modelorder} and~\ref{fig:sim_bin_overview} show \wSSE{}
performing uniformly poorly across the simulation grid.

\begingroup
\renewcommand{\arraystretch}{0.94}
\begin{longtable}{@{}l r c@{}}
\caption{Tukey--Kramer HSD connecting letters for the Gaussian
simulation (response: log-NRMSE on a noiseless holdout; smaller least-squares
means are better; rows are sorted from worst to best).}
\label{tab:tukey-hsd}\\
\toprule
\multicolumn{1}{@{}l}{Setting} &
\multicolumn{1}{r}{Least-squares mean} &
\multicolumn{1}{c@{}}{Group} \\
\midrule
\endfirsthead
\toprule
\multicolumn{1}{@{}l}{Setting} &
\multicolumn{1}{r}{Least-squares mean} &
\multicolumn{1}{c@{}}{Group} \\
\midrule
\endhead
\bottomrule
\endfoot
CV\_relaxTRUE\_lasso\_dbTRUE & -0.4550838 & A \\
CV\_relaxTRUE\_default\_dbTRUE & -0.4568168 & A \\
CV\_relaxTRUE\_lasso\_dbFALSE & -0.4784299 & B \\
CV\_relaxTRUE\_default\_dbFALSE & -0.4795369 & B \\
SVEM\_wBIC\_relaxFALSE\_lasso\_dbFALSE & -0.4872574 & BC \\
SVEM\_wBIC\_relaxFALSE\_default\_dbFALSE & -0.4873671 & BC \\
CV\_relaxFALSE\_lasso\_dbTRUE & -0.4890622 & C \\
CV\_relaxFALSE\_default\_dbTRUE & -0.4955437 & C \\
CV\_relaxFALSE\_lasso\_dbFALSE & -0.5125708 & D \\
SVEM\_wBIC\_relaxFALSE\_default\_dbTRUE & -0.5153720 & DE \\
SVEM\_wBIC\_relaxFALSE\_lasso\_dbTRUE & -0.5154061 & DE \\
SVEM\_wAIC\_relaxFALSE\_lasso\_dbTRUE & -0.5173612 & DE \\
SVEM\_wAIC\_relaxFALSE\_default\_dbTRUE & -0.5173877 & DE \\
CV\_relaxFALSE\_default\_dbFALSE & -0.5182228 & DE \\
SVEM\_wBIC\_relaxTRUE\_lasso\_dbTRUE & -0.5202504 & DE \\
SVEM\_wBIC\_relaxTRUE\_default\_dbTRUE & -0.5205100 & DE \\
SVEM\_wAIC\_relaxTRUE\_default\_dbTRUE & -0.5240749 & E \\
SVEM\_wAIC\_relaxTRUE\_lasso\_dbTRUE & -0.5243519 & E \\
SVEM\_wAIC\_relaxFALSE\_default\_dbFALSE & -0.5429558 & F \\
SVEM\_wAIC\_relaxFALSE\_lasso\_dbFALSE & -0.5436361 & F \\
SVEM\_wBIC\_relaxTRUE\_lasso\_dbFALSE & -0.5495013 & F \\
SVEM\_wBIC\_relaxTRUE\_default\_dbFALSE & -0.5501097 & F \\
SVEM\_wAIC\_relaxTRUE\_default\_dbFALSE & -0.5745804 & G \\
SVEM\_wAIC\_relaxTRUE\_lasso\_dbFALSE & -0.5746790 & G \\
\end{longtable}
\endgroup

\begin{longtable}{@{}l r c@{}}
\caption{Tukey--Kramer HSD connecting letters for the binomial
simulation (response: \texttt{LogLoss\_Holdout}; smaller least-squares means
are better; rows are sorted from worst to best).}
\label{tab:tukey-binomial}\\
\toprule
\multicolumn{1}{@{}l}{Setting} &
\multicolumn{1}{r}{Least-squares mean} &
\multicolumn{1}{c@{}}{Group} \\
\midrule
\endfirsthead
\toprule
\multicolumn{1}{@{}l}{Setting} &
\multicolumn{1}{r}{Least-squares mean} &
\multicolumn{1}{c@{}}{Group} \\
\midrule
\endhead
\bottomrule
\endlastfoot
CV\_relaxTRUE\_default         & 0.938427 & A \\
CV\_relaxTRUE\_lasso           & 0.896920 & A \\
SVEM\_wAIC\_relaxTRUE\_default & 0.808413 & B \\
SVEM\_wAIC\_relaxTRUE\_lasso   & 0.800409 & B \\
SVEM\_wBIC\_relaxTRUE\_default & 0.735577 & C \\
SVEM\_wBIC\_relaxTRUE\_lasso   & 0.724202 & C \\
SVEM\_wAIC\_relaxFALSE\_default& 0.683151 & D \\
SVEM\_wAIC\_relaxFALSE\_lasso  & 0.683002 & D \\
CV\_relaxFALSE\_lasso          & 0.602811 & E \\
CV\_relaxFALSE\_default        & 0.593384 & E \\
SVEM\_wBIC\_relaxFALSE\_default& 0.575261 & E \\
SVEM\_wBIC\_relaxFALSE\_lasso  & 0.575050 & E \\
\end{longtable}

\section*{CRediT authorship contribution statement}
Andrew T. Karl: Software; Writing - original draft.

\section*{Funding}
This research did not receive any specific grant from funding agencies in the public, commercial, or not-for-profit sectors.

\section*{Declaration of competing interest}
The author developed the \SVEMnet package and the related JMP add-in. The author has no other competing financial interests or personal relationships that could have appeared to influence the work reported in this paper.

\section*{Declaration of generative AI and AI-assisted technologies in the manuscript and software preparation process}

During the preparation of this manuscript and the accompanying software, the author used OpenAI GPT models as assistive tools. Under the author's direction, these models were used (i) to revise and proofread prose and (ii) to help draft the \SVEMnet{} (and simulation) R code and associated documentation. Claude (Anthropic) was also used to proofread this revision. All AI-generated text and code were reviewed, edited, and validated by the author before inclusion. All algorithmic design choices, statistical methodology, simulation design, and final implementation decisions were made by the author, who takes full responsibility for the content of the manuscript, the software, and all reported results.

\section*{Data and software availability}

\textbf{Data and replication.} The historical R scripts and CSV outputs
for the simulations, figures, and LNP case study are archived in version~4 of
the Mendeley Data repository \citep{Karl2025_SVEMnetData}. Version~5
(10 August 2026; \url{https://doi.org/10.17632/2zshchrhjh.5}) updates the
bundled package to 3.5.0. Version~4 retains the 3.2.0 source used for the
historical analyses.

\textbf{Software.} \SVEMnet is available on CRAN:
\url{https://doi.org/10.32614/CRAN.package.SVEMnet}. The analyses in this paper were run with \SVEMnet{} v3.2.0; the corresponding source tarball is included in the replication archive \citep{Karl2025_SVEMnetData}. The current CRAN release checked for this revision is 3.6.2
(18 September 2026); its reference manual and NEWS document the current API
and intervening corrections. The study used R 4.5.2. The package is
licensed under GPL-2 \textbar{} GPL-3.

\textbf{JMP add-in.} An optional JMP add-in,
``Relaxed Elastic Net and Lasso SVEM using R''
(\url{https://marketplace.jmp.com/appdetails/Relaxed+Elastic+Net+and+Lasso+SVEM+using+R}),
is available via the JMP Marketplace. The add-in lets JMP users
 call \SVEMnet for a saved \texttt{Fit Model} table script with Gaussian responses
and obtain prediction formula columns.

\section*{Independently tested by}
Daniel Fortune, Hexion, Inc., Katy, Texas, USA, installed \SVEMnet{} (version 3.2.0) from a local build in  R 4.5.2 and ran the supplemental \texttt{LNP\_example\_SVEMnet.R} code shown in Section~\ref{sec:example} (and available for download from the repository \citep{Karl2025_SVEMnetData}). The $p$-values printed to the console match those reported in Section~\ref{sec:example}. The CSV file of candidates saved successfully and contained the same entries (where \texttt{candidate\_type} is equal to \texttt{best}) as shown in Table~\ref{tab:lnp-cands-recipe} and Table~\ref{tab:lnp-cands-behav}.

\section*{Acknowledgements}
The author thanks Daniel Fortune for verifying installation and the core workflow, and the anonymous reviewers who provided feedback that led to improvements in the paper.

\bibliographystyle{elsarticle-num-names}
\bibliography{main_v5}
\end{document}